\documentclass{emulateapj}

\usepackage{natbib}
\shorttitle{YSOVAR: Mid-IR variability in L1688}
\shortauthors{H. M. G\"unther for the YSOVAR team}

\begin{document}
\title{YSOVAR: Mid-IR variability in the star forming region Lynds 1688}
\author{H.~M.~G\"unther}
\affil{Harvard-Smithsonian Center for Astrophysics, 60 Garden Street, Cambridge, MA 02138, USA}
\email{hguenther@cfa.harvard.edu}

\and
\author{A. M. Cody}
\affil{Spitzer Science Center, California Institute of Technology, Pasadena, CA 91125, USA}
\and
\author{K. R. Covey}
\affil{Lowell Observatory, 1400 W. Mars Hill Rd., Flagstaff, AZ 86001, USA}
\and
\author{L. A. Hillenbrand}
\affil{Department of Astronomy, California Institute of Technology, Pasadena, CA 91125, USA}
\and
\author{P. Plavchan}
\affil{NASA Exoplanet Science Institute, California Institute of Technology, 770 South Wilson Avenue, Pasadena, CA 91125, USA}
\and
\author{K. Poppenhaeger}
\affil{Harvard-Smithsonian Center for Astrophysics, 60 Garden Street, Cambridge, MA 02138, USA}
\affil{NASA Sagan fellow}
\and
\author{L. M. Rebull}
\affil{Spitzer Science Center/Caltech, 1200 E. California Blvd., Pasadena, CA 91125, USA}
\and
\author{J. R. Stauffer}
\affil{Spitzer Science Center/Caltech, 1200 E. California Blvd., Pasadena, CA 91125, USA}
\and
\author{S. J. Wolk}
\affil{Harvard-Smithsonian Center for Astrophysics, 60 Garden Street, Cambridge, MA 02138, USA}

\and
\author{L. Allen}
\affil{National Optical Astronomy Observatories, Tucson, AZ, USA}
\and
\author{A. Bayo}
\affil{Max Planck Institut f\"ur Astronomie, K\"onigstuhl 17, 69117, Heidelberg, Germany}
\affil{Departamento de F\'isica y Astronom\'ia, Facultad de Ciencias, Universidad de Valpara\'iso, Av. Gran Breta\~na 1111, 5030 Casilla, Valpara\'iso, Chile }
\and
\author{R. A. Gutermuth}
\affil{Dept. of Astronomy, University of Massachusetts, Amherst, MA  01003, USA}
\and
\author{J. L. Hora}
\affil{Harvard-Smithsonian Center for Astrophysics, 60 Garden Street, Cambridge, MA 02138, USA}
\and
\author{H. Y. A. Meng}
\affil{Infrared Processing and Analysis Center, California Institute of Technology, MC 100-22, 770 S Wilson Ave, Pasadena, CA 91125, USA}
\affil{Lunar and Planetary Laboratory, University of Arizona, 1629 E University Blvd, Tucson, AZ 85721, USA }
\and
\author{M. Morales-Calder\'on}
\affil{Centro de Astrobiolog\'ia (INTA-CSIC), ESAC Campus, P.O. Box 78, E-28691 Villanueva de la Canada, Spain}
\and
\author{J. R. Parks}
\affil{Department of Physics and Astronomy, Georgia State University, 25 Park Place South, Atlanta, GA 30303, USA}
\and
\author{Inseok. Song}
\affil{Physics and Astronomy Department, University of Georgia, Athens, GA 30602-2451, USA}

\begin{abstract}
The emission from young stellar objects (YSOs) in the mid-IR is dominated by the inner rim of their circumstellar disks. We present an IR-monitoring survey of $\sim800$ objects in the direction of the Lynds 1688 (L1688) star forming region over four visibility windows spanning 1.6~years using the \emph{Spitzer} space telescope in its warm mission phase. Among all lightcurves, 57 sources are cluster members identified based on their spectral-energy distribution and X-ray emission. Almost all cluster members show significant variability. The amplitude of the variability is larger in more embedded YSOs. Ten out of 57 cluster members have periodic variations in the lightcurves with periods typically between three and seven days, but even for those sources, significant variability in addition to the periodic signal can be seen. No period is stable over 1.6~years. Non-periodic lightcurves often still show a preferred timescale of variability which is longer for more embedded sources. About half of all sources  exhibit redder colors in a fainter state. This is compatible with time-variable absorption towards the YSO. The other half becomes bluer when fainter. These colors can only be explained with significant changes in the structure of the inner disk. No relation between mid-IR variability and stellar effective temperature or X-ray spectrum is found.
\end{abstract}


\section{Introduction}
\label{sect:introduction}

Stars form in dense and cool molecular clouds. When the local density is high enough, the matter can gravitationally collapse and form a young stellar object (YSO). In the early phases, the thick envelope dominates the emission from the YSO and hides what is going on within (class~I). Eventually, the envelope flattens out to a circumstellar accretion disk. This disk still causes an infrared (IR) excess above the level of a stellar photosphere (class~II or classical T Tauri star - CTTS), which can be used to distinguish those objects from main-sequence stars, for example using the \emph{Spitzer} space telescope \citep{2004ApJS..154....1W} InfraRed Array Camera \citep[IRAC,][]{2004ApJS..154...10F}.  When the disk is cleared, the IR colors of the YSO match those of main-sequence stars (class~III or weak-lined T Tauri star - WTTS). In addition to the circumstellar absorption, many YSOs are embedded in the molecular cloud, so that even class~III objects can appear reddened. 

The accretion disk does not reach down to the central star. Instead, the inner edge of the gas disk is truncated by the stellar magnetic field. The inner radius of the optically thick dust in the disk is larger than the inner radius of the gas disk and mostly given by the dust-sublimation temperature. Some of the mass in the circumstellar disk condenses into planets, some is blown out by accretion-driven disk and stellar winds, and is accreted onto the central star. This accretion can happen via magnetically confined accretion funnels \citep[e.g.,][]{1994ApJ...429..781S} or via some magneto-hydrodynamical instability \citep[e.g.,][]{2012MNRAS.421...63R}.

T~Tauri Stars (TTS) were originally identified by their variability \citep{1945ApJ...102..168J} -- long before anybody realized that TTS are indeed pre-main sequence stars. The dominant timescale in the optical is the stellar rotation period, typically a few days to a week or more \citep{1983ApJ...267..191R,1986A&A...165..110B,2009ApJ...694L.153N}. YSOs can have cool spots caused by magnetic activity similar to our Sun and also hot spots which mark the impact points of the accretion funnels onto the stellar surface \citep[see, e.g., review by][]{2013AN....334...67G}. This impact happens at free-fall velocities up to 500~km~s$^{-1}$; thus, the accretion shock heats the accreted mass to X-ray emitting temperatures \citep[see, e.g., reviews by][]{2004A&ARv..12...71G,
2011AN....332..448G}. In the optical, the accretion region appears as emission that often is approximated as a blackbody with temperature $T < 10\,000$~K (\citealp{calvetgullbring,2012ApJ...752L..20I}, but see also \citealp{2012AstL...38..649D,2013AstL...39..389D} who argue that line emission contributes to the veiling in addition to a continuum). Variability in the mass accretion rate can lead to changes in the hot spot signatures. 

The dynamical timescale that controls the accretion is the Keplerian period of the inner disk where the accretion funnels start. The inner disk radius is found close to the co-rotation radius leading to a typical timescale of a few hours for typical masses and rotation periods of YSOs. Indeed optical variability with amplitudes around 0.1~mag is often observed in CTTS on this timescale \citep{1996MNRAS.282..167S,2008MNRAS.391.1913R}. Another source of variability related to the accretion could be oscillations of the accretion shock on timescales of seconds. This has been predicted theoretically \citep[e.g.,][]{2008MNRAS.388..357K}, but is not observed so far \citep{2009ApJ...703.1224D,2010A&A...518A..54G}, possibly because the accretion spot separates into many small funnels that oscillate independently at different phases and frequencies. However, \citet{2011AJ....142..141B} find indications that strong accretion in V1647~Ori could excite radial pulsations of the star itself.

One of the largest classes of short timescale ($\tau < 10$ days) optical and IR variability in YSOs is that due to variable extinction events \citep{cody2014,2014arXiv1401.6600S}.
These come in three categories - AA Tau-type variables (stars with broad, periodic flux dips, whose amplitudes can be up to a magnitude or more in the optical), presumably due to our line of sight passing through a warp in the inner circumstellar disk; stars with similar or narrower flux dips that have no obvious periodicity - presumably due to stars where our line of sight passes close to the disk and where disk instabilities can levitate dust high enough above the plane to intersect our line
of sight briefly; and stars with narrow, periodic flux dips - perhaps where our line of sight is being intersected by dust entrained in material accreting onto the star in a funnel flow.  About 20\% of the YSOs in NGC~2264 fall into one of these categories in the sample of \citet{cody2014}.

YSOs can also vary on much longer timescales. Variability on the timescales of
years could be caused by changing circumstellar extinction
\citep{2007A&A...461..183G} for a Keplarian disk around a solar-mass YSO this timescale translates to a radius of a few AU) or by massive accretion events when a significant fraction of the disk mass drains onto the YSO \citep{2012Natur.487...74M}. In this case the accretion luminosity can outshine the YSO by orders of magnitudes and it takes months to years \citep[in the case of EXor outbursts,][]{2012ApJ...749..188L} or even centuries \citep[FUor outbursts,][]{1996ARA&A..34..207H} until the accretion decays back down to the original level.

In any lightcurve, several of the processes dicussed above can contibute to the observed variability at the same time and it depends on the properties of each object which one dominates and if secondary effects can be detected in the lightcurve. For example, cool spots, hot spots, absorption, and massive accretion events can all influence the same optical light curve. Another case are X-rays, where the flux and the spectrum can change due to periodic absorption \citep[AA~Tau,][]{2007A&A...462L..41S}, variability in the accretion rate \citep[TW~Hya,][]{2010ApJ...710.1835B} or coronal activity similar to what is seen on the sun. In most YSOs, the last point is dominant and X-ray lightcurves often show the fast rise in flux and temperature and a slower decay characteristic of coronal activity \citep[see, e.g., the \emph{Chandra} monitoring of the Orion Nebular Cluster,][]{2005ApJS..160..423W,2005ApJS..160..319G}.

The spectral energy distribution (SED) of YSOs in the optical is dominated by the stellar photosphere and the accretion spot. Thus, optical monitoring is very effective for understanding the stellar rotation and the accreting spot. However, the disk radiates mostly at longer wavelengths, which are probed in the IR observations presented in this article. Depending on the mass of the disk and the size of the inner hole, the disk will start to dominate the SED at the $K_s$ band or in the IRAC bands at $3.6\,\mu$m and $4.5\,\mu$m. Simple disk models still treat the disk as a static and axisymmetric structure, but observationally it now seems that the disk is in fact ``a bubbling, boiling, wrinkled, dented, warped mass of gas and dust'' \citep{2013AJ....145...66F}, see also \citet{cody2014,2014arXiv1401.6600S}.

This paper is part of the YSOVAR (Young Stellar Object VARiability) project, which has monitored the \object{Orion Nebular Cluster} (ONC) and eleven smaller star forming regions with IRAC in $3.6\,\mu$m and $4.5\,\mu$m to understand the mid-IR variability of YSOs. First results on the ONC are published in \citet{2011ApJ...733...50M,2012ApJ...753..149M}. More details of the observing strategy and an overview of the data can be found in \citet{luisa} (from now on ``paper~I''). Comparing data from all clusters, paper~I defines certain cut-off values for the data reduction, e.g.\ how much variability in a lightcurve is required to reliably identify an object as variable. In the analysis, paper~I concentrates on variability in the IR on timescales of years.
In this article, we present a \emph{Spitzer}/IRAC monitoring campaign of the star forming region L1688 in the mid-IR to characterize the variability timescales and amplitudes as well as the color changes in the mid-IR in much more detail for the objects in L1688 than paper~I on timescales up to two years.

The structure of this paper is as follows:
First, we introduce L1688, the star forming region targeted by these observations (section~\ref{sect:thestarformingregionl1688}).
In section~\ref{sect:observationsdatareductionandauxiliarydata} we introduce the data reduction and discuss source lists and stellar properties obtained from the literature. Section~\ref{sect:midirvariability} classifies all sources according to their variability. Section~\ref{sect:resultsanddiscussion} presents our results and discusses physical models to explain the observed features in the lightcurves. We end with a summary and some conclusions in section~\ref{sect:summary}.

\section{The star forming region L1688}
\label{sect:thestarformingregionl1688}

\begin{figure*}
\plotone{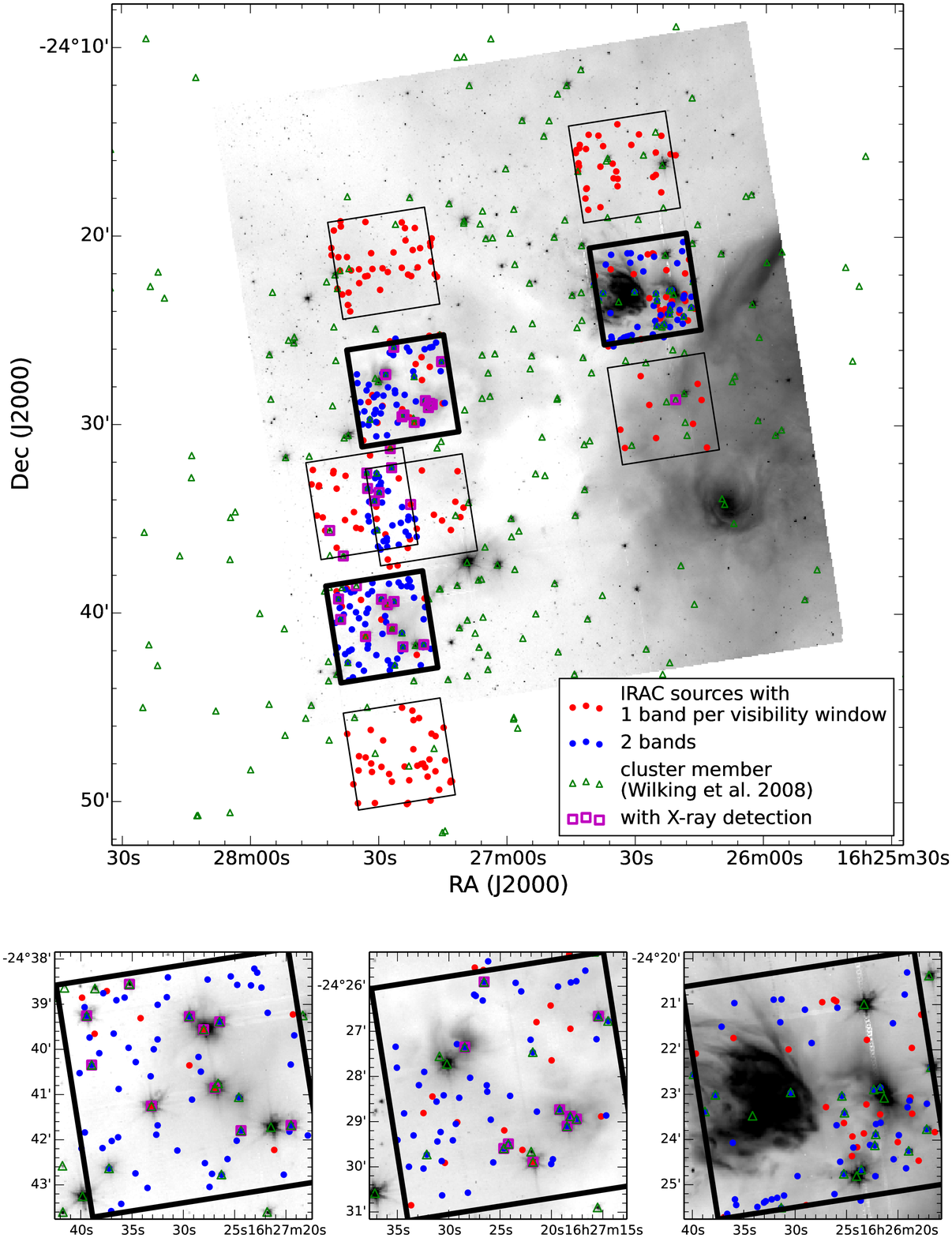}
\caption{Locations of sources with more than five datapoints in their lightcurves. 
These sources are seen in three groups of three fields. The central field of each group is the primary target field (thick black squares) and is observed in both channels in each visibility window. 
The regions to the north and south (thin black squares) of it are observed in one channel per visibility window only.
IRAC sources that are cross-matched with a \emph{Chandra} X-ray source are surrounded by a magenta square.
Additionally, known cluster members from \protect{\citet{2008hsf2.book..351W}} are marked.
The background is an inverse gray scale image in $3.6\,\mu$m obtained during the \emph{Spitzer} cryogenic mission. 
The white patch in the center of the image is the L1688 dark cloud. The bright star surrounded by a reflection nebula in the south-west is EM*~SR~3. 
\emph{lower panels}: Zoom into the three target fields.
\label{fig:map}}
\end{figure*}
Lynds~1688 (L1688) is a sub-cloud of the $\rho$~Ophiuchus star forming region, one of the best-studied young clusters in the sky \citep[see e.g.\ review by][]{2008hsf2.book..351W}.
%VLBA measurements and Hipparcos parallaxes point to values around 120~pc \citep{2008ApJ...675L..29L,2008A&A...480..785L}, while \citet{2008AN....329...10M} find 135~pc, again based on Hipparcos data. One possible solution is that the streamers of L1688 are closer to the Sun than the core of the region \citep{2008A&A...480..785L}. In this article, we adopt a distance of 120~pc, but we note that most of our results are independent of the actual distance. 
The central region of L1688 is very dense and deeply embedded ($A_V=50-100$~mag, see Figure~\ref{fig:map}). Thus, all surveys of the regions necessarily miss some cluster members. An extinction limited, spectroscopic survey \citep{2011AJ....142..140E} finds an average age of 3.1~Myr for a 6.8~pc$^2$ region centered on L1688 and no significant deviation from the initial mass function. Earlier studies \citep{1995ApJ...450..233G,1999ApJ...525..440L, 2002A&A...393..597N} concentrated on the deeper embedded core and found a much younger age of 0.3~Myr. Some, but not all, of this difference is due to the specific reddening laws or pre-main sequence evolutionary tracks used in these studies \citep[see discussion in][]{2011AJ....142..140E}.

\citet{2012PASP..124.1137F} present photometric and spectroscopic monitoring for five YSOs in L1688 nearly simultaneous with the \emph{Spitzer} observations discussed here. They do not see any correlation between the hydrogen emission lines that are usually considered accretion indicators and the features in the IR lightcurves of their targets, indicating that the relatively modest variability they observed is not caused by changes in the accretion rate. Additional notable objects with well-sampled NIR lightcurves are \object{WL 4} \citep{2008ApJ...684L..37P} and \object{YLW 16A} \citep{2013A&A...554A.110P}, which show eclipses with periods of 131 and 93~days, respectively. These sources can be interpreted as multiple systems, where one or more components are eclipsed by a warped circumstellar disk. 

Thus far, the most comprehensive study of near-IR variability of YSOs in L1688 is \citet{2013arXiv1309.5300P} (see references therein for other IR variability studies), who make use of a Two Micron All Sky Survey \citep[2MASS][]{2006AJ....131.1163S} calibration field that overlaps L1688 so each source has up to 1584 datapoints in $J$, $H$, and $K_s$ spanning 2.5~years. They find 79\% of the known YSOs to be variable. In total, 32 sources are periodic (including cool starspots, hot accretion spots and 6 systems with eclipses), 31 sources show a long-term trend and 40 sources vary aperiodically on shorter timescales. The new data presented in this article complements the \citet{2013arXiv1309.5300P} study with observations at longer wavelengths.

\section{Observations, data reduction and auxiliary data}
\label{sect:observationsdatareductionandauxiliarydata}

In this section we briefly describe the data reduction for the \emph{Spitzer} and \emph{Chandra} observations. A detailed account of the observations, data processing and the source extraction is given in paper~I.  We also give an overview of auxiliary data on the stellar properties retrieved from the literature, which we need to test if the variability characteristics depend on the central star. We then assess the cluster membership and SED class for every source with a usable lightcurve.

\subsection{Spitzer}
\label{sect:spitzer}

\subsubsection{Spitzer observations}
\label{sect:spitzerobservations}

\begin{deluxetable}{ccc}
\tablecaption{\label{tab:aor} Observation Log.}
\tablehead{\colhead{AOR} & \colhead{Start time} & \colhead{End time}}
\startdata
29267200 & 2010-04-12 11:00:41 & 2010-04-12 11:18:44 \\
29266688 & 2010-04-12 17:02:37 & 2010-04-12 17:28:01 \\
29266176 & 2010-04-12 22:48:49 & 2010-04-12 23:09:28 \\
29265664 & 2010-04-13 08:44:31 & 2010-04-13 09:05:35 \\
29265408 & 2010-04-13 20:09:17 & 2010-04-13 20:26:33 \\
\enddata
\tablecomments{This table is published in its entirety in the electronic edition of this journal. A portion is shown here for guidance regarding its form and content.}
\end{deluxetable}

\begin{figure}
\plotone{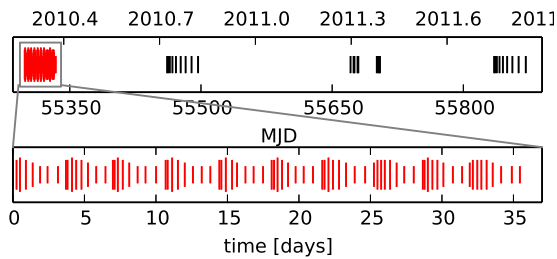}
\caption{Temporal spacing of observations of L1688. The ``fast cadence'' observations in the first visibility window are shown in red. 
Each observation is marked by a line, whose length is proportional to the number of observations in a 24 hour window centered on that observation
so that the frequency of observations can be judged where symbols overlap.
\label{fig:cadence}}
\end{figure}

Three fields in L1688 were observed with \emph{Spitzer} in four observing windows from 2010-04-12 to 2010-05-16 (visibility window 1), 2010-09-22 to 2010-10-27 (visibility window 2), 2011-04-20 to 2011-05-23 (visibility window 3), and 2011-10-01 to 2011-11-06 (visibility window 4). These windows are consecutive visibility periods dictated by the \emph{Spitzer} orbit \citep{2004ApJS..154....1W}. They are shown visually in Figure~\ref{fig:cadence}. In the first visibility window, the sampling is much denser in time than in the later visibility windows. For most sources, about 70 observations with irregular time intervals to aid period detection were obtained in visibility window 1. In the first visibility window, we use a repeating pattern of 8 observations every 3.5 days. Within the 3.5 day period the time-step increases from 2 to 16 hours. In the last three visibility windows, the time steps increase linearly with step lengths of roughly 1, 2, 3, ... days to again sample multiple variability frequencies equally.  This irregular sampling minimizes aliasing with the observation frequency.
Less than ten datapoints per source are taken in each of the later three visibility windows. In total, there are 108 observations with a total mapping time of 30.7 hours.  Table~\ref{tab:aor} lists the time of each observation. They can be found under Program Identification number (PID) 61024 in the Spitzer Heritage Archive. Each observation consists of six dithers in IRAC mapping mode using the high-dynamic-range (HDR) data acquisition mode which obtains a 0.4~s and a 10.4~s exposure for each pointing. 

The three fields chosen were observed with the IRAC~1 and IRAC~2 channels (effective wavelengths 3.6$\,\mu$m and 4.5$\,\mu$m). Both channels operate simultaneously, but their fields-of-view are non-overlapping. Thus, each target field is observed in two consecutive pointings, one for each channel. A secondary field is observed in the secondary channel  while the primary channel is observing the target field. In visibility windows 1 and 3, sources South of the main fields have only IRAC~1 data, while those to the North only have IRAC~2 data. In visibility windows 2 and 4, the situation is reversed. Not all sources in the central fields have usable data in both bands, because they might be too bright or too faint in one channel, or fall on the edge of the map. Additional sources with two band coverage are found where the northern side field of one target field overlaps with the southern side field of another target field.

\subsubsection{Spitzer data reduction}
\label{sect:spitzerdatareduction}

Here we summarize the main data reduction and processing steps described in detail in paper~I. Basic calibrated data (BCD) are obtained from the \emph{Spitzer} archive. Further data reduction is performed with the IDL package cluster grinder \citep{2009ApJS..184...18G}, that treats each BCD image for bright source artifacts. Aperture photometry is performed on individual BCDs with an aperture radius of $2\farcs4$. To increase the signal-to-noise ratio and to reject cosmic rays, the photometry from all BCDs in each observation is combined. The reported value is the average brightness of all BCDs within that observation that contain the source in question, after rejecting outliers. The photometric uncertainties obtained from the aperture photometry are, particularly for faint sources, only lower limits to the total uncertainty, since distributed nebulosity often found in star forming regions can contribute to the noise. To improve these estimates, paper~I introduces an error floor value that is added in quadrature to the uncertainties of individual photometric points. The value of the error floor is 0.01~mag for IRAC~1 and 0.007~mag for IRAC~2.

We cross-match sources from individual observations with a matching radius of $1\arcsec$ with each other and with the 2MASS catalog, which is used as a coordinate reference. All photometric measurements performed in the context of the YSOVAR project are collected in a central database, which we intend to deliver to the Infrared Science Archive (IRSA) for general distribution. 
Data for this article were retrieved from the YSOVAR database on 2013-10-31 and further processed using custom routines in Python available at \url{https://github.com/YSOVAR}. 
We visually checked all frames for lightcurves that are classified as variable in section~\ref{sect:midirvariability} and removed datapoints visibly affected by instrumental artifacts (cosmic rays, read-out streaks for bright neighbors). Figure~\ref{fig:map} shows the distribution of the sources with lightcurves in our input catalog overlayed on a larger IRAC~1 map observed during the cryogenic mission.

In this article, we consider only objects that have at least five datapoints in our IRAC~1 
or IRAC~2 lightcurves. 
A stricter definition is employed in paper~I, where only sources with more than five datapoints 
\emph{in the fast-cadence data (first visibility window)} are used. In L1688, 822 of the total 
list of 882 sources fullfill this stricter condition. In
Table~\ref{tab:tab2} they are marked in the column \texttt{StandardSet}.

\subsubsection{Instrumental effects remaining in IRAC lightcurves}
\label{sect:instrumentaleffectsremaininginiraclightcurves}

\begin{figure}
\plotone{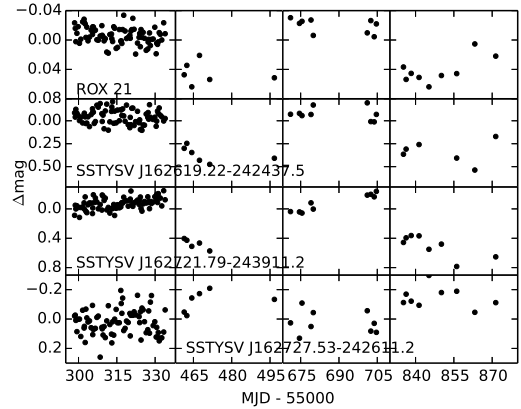}
\caption{Lightcurves with apparent jumps between visibility windows. The examples shown here are all $4.5\,\mu$m lightcurves, but the same can be seen in the $3.6\,\mu$m channel for
some sources. 
The mean brightness of these objects is (from top to bottom) 8.2, 14.2, 14.6 and 14.8~mag.
\label{fig:instrumentallcs}}
\end{figure}

Despite the careful data reduction described above, some residual artifacts remain in the \emph{Spitzer} lightcurves. In this section, we search for artifacts that are related to the position on the detector. Compared with other clusters in the YSOVAR project, L1688 is particularly suited to discover these kinds of effects because the fields observed in L1688 have almost no rotation within one visibility window. The spacecraft orientation flips between visibility windows, so that most instrumental artifacts produce lightcurves that have one level in visibility window 1 and 3 and another level in visibility window 2 and 4. Figure~\ref{fig:instrumentallcs} shows sources with a magnitude between 8 and 15 in $4.5\,\mu$m, that might fall in this category. The relative difference in magnitude is smaller for brighter sources, but since the photometric errors are also smaller these instrumental effects can still be significant. 

We visually inspected every frame for a sample of sources with lightcurves similar to those in Figure~\ref{fig:instrumentallcs} to identify the cause for the artifacts. We found that several different effects can cause these steps in the lightcurve. Some sources are close to a detector edge, such that the background is not well-determined, some are in the wings of the point-spread function (PSF) of a bright neighbor, and some show residuals from hot pixels. Proximity to a detector edge or hot pixel affects only one of the spacecraft orientations; even the PSF wings change with the orientation since the PSF is not circular. 

Sufficiently strong intrinsic variability can mask this offset between visibility windows and for any individual source, this instrumental effect cannot be distinguished from intrinsic variability with a period of one year. Thus, we use a statistical approach to quantify the number of lightcurves that suffer from this problem. For each channel, we have about 200 lightcurves with datapoints in all four observing visibility windows (the total number of lightcurves is larger, but only for sources in the three target fields do we have data for all four visibility windows). If the mean magnitude in a visibility window depends on the detector position of the source, this source will on average be brighter in visibility window 1 and 3 and fainter in visibility window 2 and 4 (or vice-versa). We calculate the mean magnitude in each visibility window and test whether the two brightest mean magnitudes belong to visibility window [1,3] or [2,4] or any other combination ([1,2], [1,4], [2,3], [3,4]). 

We use the following abbreviated notation: When we calculate the mean magnitudes in each visibility window for a sample of $n=a+b+c$ sources, then $a$:$b$:$c$ means that $a$ sources have the brightest mean magnitudes in visibility windows [1,3], $b$ sources in visibility windows [2,4] and $c$ sources have their two brightest mean magnitudes in any other combination of visibility windows. If the difference in the mean magnitude between visibility windows is unrelated to the position on the detector, we expect the ratio $\frac{1}{6}:\frac{1}{6}:\frac{2}{3} = $17\%:17\%:67\%. In contrast, the observed lightcurves have the ratio 47:66:75=25\%:35\%:40\% for IRAC~1 and 52:59:95=25\%:29\%:46\% for IRAC~2. This is incompatible with the expected multinomial distribution (the probability to observe a distribution at least as far from the expected $\frac{1}{6}:\frac{1}{6}:\frac{2}{3}$ by chance is $<10^{-8}$). In each case, about 50 lightcurves, a quarter of the sample, need to be shifted from the first two bins to the last bin to make the observed distribution compatible with the expected distribution. This implies that about a quarter of all lightcurves suffer from the artifacts described above. However, both IRAC channels use independent detectors and thus the chance that both channels are affected for the same source and that the effect goes the same way (bright-faint, vs. faint-bright) is low.

Below in section~\ref{sect:midirvariability} we identify variable sources using a Stetson and a $\chi^2$ test. The limits in those tests are designed to be conservative and indeed we find a distribution of 5:7:24=14\%:19\%:67\% for IRAC~1 and 5:5:27=14\%:14\%:73\% for IRAC~2, when we restrict the sample to those lightcurves that will be classified as variable below. Both ratios are fully compatible with the expected multinomial distribution $\frac{1}{6}:\frac{1}{6}:\frac{2}{3}$. This shows that the limits we apply are conservative enough that the sample of stars we identify as variable has no or only few sources where the variability is not due to intrinsic source variability. Therefore it is not necessary to remove any source based on magnitude jumps between visibility windows.  Only one of the four lightcurves shown in Figure~\ref{fig:instrumentallcs} (SSTYSV J162727.53-242611.2) will be identified as variable below. We visually inspected all lightcurves that are marked as variable and, apart form the example shown above, we did not see lightcurves where the variability seems to be due to the pattern discussed in this section.

In summary, about a quarter of all lightcurves are affected by detector position dependent artifacts. In extreme cases, the associated jumps reach 0.5~mag for a 14~mag source, and up to 0.05~mag in bright sources around 8~mag. However, we show statistically that the definition of variability we use is so conservative that this effect does not contribute a significant number of objects to our sample of variable sources.

\subsection{Chandra}
\label{sect:chandra}

Disk-bearing YSOs can be identified from \emph{Spitzer} data alone, but information in other spectral bands is required to find the other cluster members.
X-ray observations are one way to identify diskless (class~III) YSOs. In the IR the SEDs of those sources are indistinguishable from a main-sequence field star, but due to their rapid rotation, YSOs are much brighter in X-rays than field stars \citep{1999ARA&A..37..363F}.

L1688 was observed by \emph{Chandra} on 2000-04-13 for 100~ks exposure time in the \texttt{FAINT} mode with the ACIS instrument (\dataset[ADS/Sa.CXO#obs/00635]{ObsId 635}). We reprocessed this exposure with the ANCHORS pipeline \citep{2007HiA....14..587W} using a recent calibration; see discussion in paper~I. These data has been analyzed in detail to study the distribution of X-ray properties in CTTS and to identify brown dwarfs in this star forming region \citep{2001ApJ...557..747I,2001ApJ...563..361I}.

During the observation, five ACIS chips were operational, four from the central ACIS-I imaging array, as well as one ACIS-S chip. The point-spread-function (PSF) degrades significantly for sources located off-axis, and thus the coordinates of the outer sources are less reliable. To cross-match X-ray sources with \emph{Spitzer} sources we used a matching radius of 1\arcsec{} for X-ray sources within 3\arcmin{} of the optical axis of \emph{Chandra}, 1.5\arcsec{} for sources between 3\arcmin{} and 6\arcmin{} away from the optical axis and 2\arcsec{} for all sources located further than 6\arcmin{} from the optical axis. The observed \emph{Chandra} field overlaps about two thirds of the area covered in the \emph{Spitzer} monitoring. However, the variable PSF leads to a sensitivity that varies over the observed field, thus the absence of an X-ray detection for \emph{Spitzer} sources does not necessarily imply the absence of X-ray emission.
The ACIS detector has an intrinsic energy resolution, and we use the net flux to characterize the X-ray properties. For sources with more than 20 counts, we also fit an absorbed single-temperature APEC model \citep{2012ApJ...756..128F} with abundances fixed at 0.3 times the solar value from \citet{1989GeCoA..53..197A}. 

% The best-fit temperature k$T$ (in keV) and column density $N_H$ (in $10^{22}$cm$^{-2}$) are given in Table~\ref{tab:tab2}.

Sources are extracted down to a very low significance. In total, there are 315 detected X-ray sources, but only 31 of them match an object with a \emph{Spitzer} lightcurve. We disregard all unmatched sources; for sources with a lightcurve and an X-ray counterpart, the X-ray properties are given in Table~\ref{tab:tab2}. To estimate the number of spurious matches, we multiply the fraction of the total survey area that is included in the positional error circles of the X-ray sources with the number of \emph{Spitzer} sources with lightcurves. The result is the average number of spurious matches. We expect at most 2-4 \emph{Spitzer} sources to be matched to a spurious X-ray source. X-ray sources that are cross-matched successfully are marked in Figure~\ref{fig:map}.

\subsection{Auxiliary data from the literature}
\label{sect:auxiliarydatafromtheliterature}

The star forming region L1688 has been the target of intense study over the past decades and a wealth of additional information exists in the literature. In particular, we refer the reader to two reviews \citep{1992lmsf.book..159W,2008hsf2.book..351W}. The latter review compiles a list of objects with a high probability of membership from a variety of published sources. The membership criteria employed are (i)~X-ray emission, which --at the distance of L1688-- is detectable only from young, and thus active stars; (ii)~optical spectroscopy, with H$\alpha$ in emission or Li in absorption; (iii)~a location above the main-sequence in the HR diagram; or (iv)~IR emission that is indicative of a circumstellar disk. 
% More recently, the region was surveyed by \citet{2012ApJ...751...22B} in $JHK_s$, to a level deeper than 2MASS \citep{2006AJ....131.1163S}.

L1688 was also observed with \emph{Spitzer} in the cryogenic mission phase with all four IRAC channels and the 24$\;\mu$m channel of the Multiband Imaging Photometer for Spitzer \citep[MIPS,][]{2004ApJS..154...25R}. Objects classified as YSOs from these data \citep{2008ApJ...672.1013P} are already contained in the membership list of \citet{2008hsf2.book..351W}. We augment our own \emph{Spitzer} data reduction with values from the catalog published by the c2d project \citep[c2d = ``From Cores to Disks'';][]{2003PASP..115..965E}. If we did not obtain a photometric value for an IRAC band or the $24\,\mu$m MIPS, but a value with the quality specifier A, B or C is present in the c2d catalog then we use that value. The data are given in Table~\ref{tab:tab2}, which specifies if a datapoint is taken from our own data reduction \citep[``G09'': using the pipeline from][]{2009ApJS..184...18G}\footnote{a subsection of the data processed with this pipeline is already presented in G09} or the c2d database.

Near-IR data is taken from 2MASS \citep{2006AJ....131.1163S} and cross-matched by the cluster grinder pipeline \citep{2009ApJS..184...18G}. Additionally, we take detections from the UKIRT Infrared Deep Sky Survey (UKIDSS) Galactic cluster survey, data release 9.
The UKIDSS project is defined in \citet{2007MNRAS.379.1599L}. UKIDSS uses the United Kingdom Infrared Telescope (UKIRT) Wide Field Camera \citep[WFCAM;][]{2007A&A...467..777C} and a photometric system described in \citet{2006MNRAS.367..454H}. The pipeline processing and science archive are described in \citet{2008MNRAS.384..637H}. We only retain detections with a \texttt{mergedClass} flag between -3 and 0. Sources brighter than $m_K=10$~mag can be saturated and we discard the UKIDSS data for any source that lies above this threshold either in UKIDSS or 2MASS. This leaves us with 2MASS data in the range $m_K=5-16$~mag and UKIDSS $m_K=10-19$~mag. The luminosity function for both dataset is almost identical from $m_K=10$ to $m_K=15$. For fainter sources, 2MASS is incomplete; UKIDSS is incomplete for $m_K > 17$~mag.

The YSOVAR data is also cross-matched with data from the SIMBAD service to provide an identification with known objects from the literature. %the NOMAD catalog \citep{2004AAS...205.4815Z}, 

In all cases the matching radius is set to $1\arcsec$. If a catalog contains multiple entries within $1\arcsec$ of a YSOVAR source, we match it to the closest catalog entry. In some cases the best cross-match is not obvious. Those sources are discussed in appendix~\ref{sect:remarksaboutcrossmatchingindividualsources}.

UKIDSS has a better spatial resolution than our IRAC data. There are nine sources where more than one UKIDSS source is found within the size of the aperture we use for IRAC photometry. In six cases (WSB 52, ISO-Oph 152, ISO-Oph 131, SSTYSV J162728.13-243719.6, SSTYSV J162718.11-244814.1, SSTYSV J162718.25-244955.8) the second source is visible in the $K$ band, the UKIDSS band that is closest in wavelength to the IRAC data, so it is likely that both sources contribute to the observed IRAC emission. In the remaining three cases (ISO-Oph 28, ISO-Oph 57, SSTYSV J162741.14-242038.3) the second source is not visible at $K$ band.

\subsection{Table of source properties}
\label{sect:tableofsourceproperties}

\begin{deluxetable*}{ccccc}
\tablecaption{\label{tab:tab2} Source designations, flux densities and lightcurve properties.}
\tablehead{\colhead{ID} & \colhead{Name} & \colhead{Unit} & \colhead{Channel} & \colhead{Comment}}
\startdata
1 & RA & deg & -- & J2000.0 Right ascension \\
2 & DEC & deg & -- & J2000.0 Declination \\
3 & name & -- & -- & identifier for object \\
4 & IAU\_NAME & -- & -- & J2000.0 IAU designation within the YSOVAR program \\
5 & other\_names & -- & -- & alternative identifiers for object \\
6 & c2d\_id &  & -- & -- \\
7 & wil08\_ID & -- & -- & Wilking et al. (2008) \\
8 & AdOC08\_AOC & -- & -- & J2000.0 IAU designation (JHHMMSS.ss+DDMMSS.s) \\
9 & UKIDSS\_sourceID & -- & -- & -- \\
10 & Jmag & mag & $J$ & -- \\
11 & e\_Jmag & mag & $J$ & observational uncertainty \\
12 & r\_Jmag & -- & $J$ & data source \\
13 & Hmag & mag & $H$ & -- \\
14 & e\_Hmag & mag & $H$ & observational uncertainty \\
15 & r\_Hmag & -- & $H$ & data source \\
16 & Kmag & mag & $K$ & -- \\
17 & e\_Kmag & mag & $K$ & observational uncertainty \\
18 & r\_Kmag & -- & $K$ & data source \\
19 & 3.6mag & mag & $3.6\;\mu$m & -- \\
20 & e\_3.6mag & mag & $3.6\;\mu$m & observational uncertainty \\
21 & r\_3.6mag & -- & $3.6\;\mu$m & data source \\
22 & 4.5mag & mag & $4.5\;\mu$m & -- \\
23 & e\_4.5mag & mag & $4.5\;\mu$m & observational uncertainty \\
24 & r\_4.5mag & -- & $4.5\;\mu$m & data source \\
25 & 5.8mag & mag & $5.8\;\mu$m & -- \\
26 & e\_5.8mag & mag & $5.8\;\mu$m & observational uncertainty \\
27 & r\_5.8mag & -- & $5.8\;\mu$m & data source \\
28 & 8.0mag & mag & $8.0\;\mu$m & -- \\
29 & e\_8.0mag & mag & $8.0\;\mu$m & observational uncertainty \\
30 & r\_8.0mag & -- & $8.0\;\mu$m & data source \\
31 & 24mag & mag & $24\;\mu$m & -- \\
32 & e\_24mag & mag & $24\;\mu$m & observational uncertainty \\
33 & r\_24mag & -- & $24\;\mu$m & data source \\
34 & SEDclass & -- & -- & IR class according to SED slope \\
35 & s1\_SEDclass & -- & -- & IR class according to SED slope (visibility window 1) \\
36 & member(YSOVAR) & -- & -- & Cluster membership according to YSOVAR standard \\
37 & StandardSet & -- & -- & Source in YSOVAR standard set? \\
38 & ns1\_36 & ct & $3.6\;\mu$m & Number of datapoints (visibility window 1) \\
39 & ns1\_45 & ct & $4.5\;\mu$m & Number of datapoints (visibility window 1) \\
40 & maxs1\_36 & mag & $3.6\;\mu$m & maximum magnitude in lightcurve (visibility window 1) \\
41 & mins1\_36 & mag & $3.6\;\mu$m & minimum magnitude in lightcurve (visibility window 1) \\
42 & maxs1\_45 & mag & $4.5\;\mu$m & maximum magnitude in lightcurve (visibility window 1) \\
43 & mins1\_45 & mag & $4.5\;\mu$m & minimum magnitude in lightcurve (visibility window 1) \\
44 & means1\_36 & mag & $3.6\;\mu$m & mean magnitude (visibility window 1) \\
45 & stddevs1\_36 & mag & $3.6\;\mu$m & standard deviation calculated from non-biased variance (visibility window 1) \\
46 & deltas1\_36 & mag & $3.6\;\mu$m & width of distribution from 10\% to 90\% (visibility window 1) \\
47 & means1\_45 & mag & $4.5\;\mu$m & mean magnitude (visibility window 1) \\
48 & stddevs1\_45 & mag & $4.5\;\mu$m & standard deviation calculated from non-biased variance (visibility window 1) \\
49 & deltas1\_45 & mag & $4.5\;\mu$m & width of distribution from 10\% to 90\% (visibility window 1) \\
50 & redchi2tomeans1\_36 & -- & $3.6\;\mu$m & reduced $\chi^2$ to mean (visibility window 1) \\
51 & redchi2tomeans1\_45 & -- & $4.5\;\mu$m & reduced $\chi^2$ to mean (visibility window 1) \\
52 & coherence\_time\_36 & d & $3.6\;\mu$m & decay time of ACF (visibility window 1) \\
53 & coherence\_time\_45 & d & $4.5\;\mu$m & decay time of ACF (visibility window 1) \\
54 & s1\_stetson\_36\_45 & -- & $3.6\;\mu$m, $4.5\;\mu$m & Stetson index for a two-band lightcurve. (visibility window 1) \\
55 & s1\_cmd\_alpha\_36\_45 & rad & $3.6\;\mu$m, $4.5\;\mu$m & angle of best-fit line in CMD (visibility window 1) \\
56 & s1\_cmd\_alpha\_error\_36\_45 & rad & $3.6\;\mu$m, $4.5\;\mu$m & uncertainty on angle (visibility window 1) \\
57 & n\_36 & ct & $3.6\;\mu$m & Number of datapoints \\
58 & n\_45 & ct & $4.5\;\mu$m & Number of datapoints \\
59 & max\_36 & mag & $3.6\;\mu$m & maximum magnitude in lightcurve \\
60 & min\_36 & mag & $3.6\;\mu$m & minimum magnitude in lightcurve \\
61 & max\_45 & mag & $4.5\;\mu$m & maximum magnitude in lightcurve \\
62 & min\_45 & mag & $4.5\;\mu$m & minimum magnitude in lightcurve \\
63 & mean\_36 & mag & $3.6\;\mu$m & mean magnitude \\
64 & stddev\_36 & mag & $3.6\;\mu$m & standard deviation calculated from non-biased variance \\
65 & delta\_36 & mag & $3.6\;\mu$m & width of distribution from 10\% to 90\% \\
66 & mean\_45 & mag & $4.5\;\mu$m & mean magnitude \\
67 & stddev\_45 & mag & $4.5\;\mu$m & standard deviation calculated from non-biased variance \\
68 & delta\_45 & mag & $4.5\;\mu$m & width of distribution from 10\% to 90\% \\
69 & redchi2tomean\_36 & -- & $3.6\;\mu$m & reduced $\chi^2$ to mean \\
70 & redchi2tomean\_45 & -- & $4.5\;\mu$m & reduced $\chi^2$ to mean \\
71 & stetson\_36\_45 & -- & $3.6\;\mu$m, $4.5\;\mu$m & Stetson index for a two-band lightcurve. \\
72 & cmd\_alpha\_36\_45 & rad & $3.6\;\mu$m, $4.5\;\mu$m & angle of best-fit line in CMD \\
73 & cmd\_alpha\_error\_36\_45 & rad & $3.6\;\mu$m, $4.5\;\mu$m & uncertainty on angle \\
74 & Teff & K & -- & effective temperature from literature \\
75 & r\_Teff & -- & -- & reference for Teff \\
76 & X-ray & -- & -- & Chandra counterpart ? \\
\enddata
\tablecomments{This table is published in its entirety in the electronic verion of the journal. Here the table columns are described as a guide to form and content.}
\end{deluxetable*}

Table~\ref{tab:tab2} contains the position, the designation, the flux densities of each source and properties of their lightcurves. The properties of the lightcurve will be discussed in detail in the remainder of this article.
Most properties of the lightcurve, e.g.\ mean, minimum and maximum, appear twice. 
They are calculated once over the entire available lightcurve and once for the first visibility window only; 
the fast-cadence sampling is available uniformly for all clusters in the YSOVAR project and thus values calculated
over the fast-cadence only can be compared between clusters (see, e.g.\ paper~I).
A subset of the properties of lightcurves with a mean magnitude $<15$ in IRAC~1 or IRAC~2 is shown in Table~\ref{tab:tab2small}.
\begin{deluxetable*}{cccccccc}
\tablecaption{Selected properties for lightcurves of variable sources with a mean magnitude $<15$ in IRAC~1 or IRAC~2 \label{tab:tab2small}}
\tablehead{\colhead{name} & \colhead{SEDclass} & \colhead{member(YSOVAR)} & \colhead{delta\_36} & \colhead{delta\_45} & \colhead{redchi2tomean\_36} & \colhead{stetson\_36\_45} & \colhead{coherence\_time\_36}}
\startdata
CFHTWIR-Oph 29 & F & yes & 0.57 & 0.55 & 368.58 & 18.61 & 3.20 \\
{[}EDJ2009] 809 & II & yes & 0.17 & 0.18 & 41.13 & --- & --- \\
WL 6 & I & yes & 0.58 & --- & 440.74 & --- & 4.70 \\
CFHTWIR-Oph 16 & II & yes & 0.05 & 0.05 & 2.42 & --- & --- \\
ISO-Oph 138 & II & yes & 0.11 & 0.12 & 15.90 & --- & 1.00 \\
ISO-Oph 53 & II & yes & 0.06 & 0.08 & 5.56 & 1.72 & 1.10 \\
WSB 52 & II & yes & 0.20 & 0.24 & 69.45 & 7.81 & 7.80 \\
WL 4 & II & yes & 0.46 & 0.38 & 366.61 & 17.97 & 5.80 \\
ISO-Oph 137 & I & yes & 0.20 & 0.17 & 65.12 & 5.60 & 1.40 \\
WL 3 & I & yes & 0.23 & 0.25 & 61.16 & 8.67 & 8.50 \\
ISO-Oph 139 & F & yes & 0.12 & 0.14 & 17.48 & 4.10 & 1.60 \\
ISO-Oph 51 & F & no & 0.40 & 0.30 & 239.84 & --- & 4.20 \\
ISO-Oph 122 & F & yes & 0.19 & --- & 45.78 & --- & 1.10 \\
WSB 49 & II & no & 0.23 & 0.18 & 84.90 & --- & --- \\
ISO-Oph 161 & I & yes & 0.34 & 0.34 & 241.40 & 13.16 & 5.40 \\
ROX 25 & II & yes & --- & 0.20 & --- & --- & --- \\
ISO-Oph 140 & II & yes & 0.17 & 0.19 & 42.95 & 5.99 & 4.70 \\
ISO-Oph 120 & F & yes & 0.18 & 0.24 & 69.78 & 6.44 & 2.20 \\
SSTYSV J162636.08-242404.2 & I & no & 0.09 & 0.07 & 6.14 & 1.28 & 0.90 \\
ISO-Oph 152 & II & yes & 0.08 & 0.07 & 5.45 & 0.73 & 1.00 \\
ISO-Oph 21 & I & yes & 0.52 & 0.35 & 329.65 & 13.48 & 7.70 \\
ROXN 44 & II & no & 0.06 & 0.05 & 5.13 & 1.86 & 3.20 \\
YLW 15 & I & yes & 0.13 & --- & 34.38 & --- & 7.20 \\
{[}GY92] 30 & I & yes & 0.13 & 0.10 & 12.90 & 2.03 & 3.60 \\
SSTYSV J162721.82-241842.4 & II & no & 0.03 & 0.06 & 0.76 & --- & 1.30 \\
WL 11 & II & yes & 0.21 & 0.25 & 28.38 & --- & --- \\
YLW 47 & II & yes & 0.16 & 0.13 & 49.72 & --- & 3.10 \\
ISO-Oph 35 & II & yes & 0.06 & 0.06 & 3.68 & --- & --- \\
{[}GY92] 264 & II & yes & 0.20 & 0.26 & 62.16 & 7.17 & 2.10 \\
2MASS J16271881-2448523 & III & no & 0.10 & 0.09 & 4.79 & --- & --- \\
ISO-Oph 153 & II & no & 0.29 & 0.23 & 118.17 & 8.42 & 0.70 \\
SSTYSV J162622.19-242352.2 & III & no & 0.02 & 0.02 & 0.68 & 0.32 & 0.70 \\
CFHTWIR-Oph 74 & II & no & 0.21 & 0.38 & 3.56 & 1.08 & 5.00 \\
ISO-Oph 34 & F & yes & 0.05 & 0.05 & 2.26 & 1.03 & 0.50 \\
CRBR 2322.3-1143 & II & yes & 0.07 & 0.12 & 4.53 & --- & 2.00 \\
ISO-Oph 33 & F & yes & 0.10 & 0.12 & 6.25 & 1.90 & 4.70 \\
ISO-Oph 145 & F & yes & 0.36 & 0.41 & 216.30 & 13.03 & 1.80 \\
ROXN 41 & II & no & 0.05 & 0.03 & 3.02 & 1.04 & 1.30 \\
ISO-Oph 144 & F & yes & 0.13 & 0.08 & 21.70 & 3.32 & 1.50 \\
SSTYSV J162617.46-242314.3 & II & no & 0.35 & 0.11 & 21.64 & 6.37 & --- \\
ISO-Oph 50 & I & yes & 0.99 & 1.90 & 821.89 & --- & 5.60 \\
{[}GMM2009] Oph L1688 30 & I & yes & 0.24 & 0.20 & 42.64 & 5.27 & 3.70 \\
ISO-Oph 165 & I & yes & 0.24 & 0.26 & 73.32 & 6.23 & 6.90 \\
{[}EDJ2009] 892 & F & yes & 0.95 & 1.13 & 1084.07 & 33.56 & 7.60 \\
ISO-Oph 26 & F & yes & 0.11 & 0.12 & 16.35 & 2.65 & 3.00 \\
ISO-Oph 154 & II & yes & 0.15 & 0.24 & 22.39 & 4.74 & 1.30 \\
CFHTWIR-Oph 21 & F & no & 0.06 & 0.05 & 5.10 & --- & 3.00 \\
ISO-Oph 124 & I & yes & 0.17 & 0.19 & 33.31 & 4.76 & 1.20 \\
WL 13 & II & yes & 0.11 & --- & 14.34 & --- & 1.70 \\
ROX 26 & I & yes & 0.29 & --- & 120.65 & --- & 8.40 \\
ISO-Oph 37 & I & yes & 0.46 & 0.36 & 228.62 & 11.54 & 6.40 \\
{[}EDJ2009] 824 & F & yes & 0.08 & 0.07 & 5.94 & 2.19 & 3.70 \\
ISO-Oph 118 & F & yes & 0.27 & 0.29 & 52.56 & 7.06 & --- \\
SSTYSV J162727.53-242611.2 & I & no & 0.31 & 0.27 & 5.03 & 1.46 & 1.10 \\
ISO-Oph 52 & F & yes & 0.30 & 0.27 & 110.30 & 9.03 & 4.90 \\
SSTYSV J162621.66-241820.1 & F & no & 0.23 & 0.28 & 7.15 & --- & 0.70 \\
ISO-Oph 19 & II & yes & 0.09 & 0.12 & 13.24 & 3.00 & --- \\
2MASS J16263046-2422571 & F & yes & 0.40 & 0.44 & 185.94 & 12.69 & 4.20 \\
SSTYSV J162728.30-244029.5 & I & no & 0.52 & 0.54 & 4.72 & 1.97 & 6.20 \\
\enddata
\tablecomments{See Table~\protect{\ref{tab:tab2}} for a detailed specification of the columns. The qualifiers 36 and 45 in the column names indicate the IRAC channel at $3.6\;\mu$m and $4.5\;\mu$m that is represented by this column.}
\end{deluxetable*}

\subsection{L1688 Membership}
\label{sect:l1688membership}

We build two L1688 membership lists based on different criteria. The first is defined in paper~I and is applied uniformly for all clusters in the YSOVAR project. Sources are treated as cluster members if they fulfill at least one of the following criteria: (i) they are classified as YSOs by \citet{2009ApJS..184...18G} based on their IR excess in cryogenic mission \emph{Spitzer} data or (ii)  they are detected as X-ray sources in \emph{Chandra} imaging and have a spectral slope compatible with a stellar photosphere (SED class~III, see section~\ref{sect:spectralslope}). At the distance of L1688 (we use 120~pc from \citet{2008A&A...480..785L}, but see also discussion in paper~I), cluster members that are young and thus still magnetically active stars can be detected in X-rays. A total of 57 sources fulfill one or both conditions. In Table~\ref{tab:tab2}, these objects are marked as ``member (YSOVAR)''. The main biases in this sample are that the IR criterion selects only those members with disks and not class~III sources, while the \emph{Chandra} criterion suffers from incomplete spatial coverage and it may include late-type foreground stars. The different biases are a common problem in multiwavelength studies of star forming regions \citep[see, e.g.,][]{2011ApJS..194...10P,2011ApJS..194...12W,2013ApJS..209...26F,2013ApJS..209...30N}.

The second membership list is taken from \citet{2008hsf2.book..351W}. On this list, 74 of our 884 sources with \emph{Spitzer} lightcurves are cluster members; 51 of those 74 are also included in the YSOVAR membership list due to their X-ray or IR emission. When we compare properties of members and non-members below without referring to a specific set, then we mean membership as defined by the standard YSOVAR criteria.

Six sources in the YSOVAR standard member set are not part of the \citet{2008hsf2.book..351W} list. More details on those source are given in Appendix~\ref{sect:g09notwil}. On the other hand, all of the sources in \citet{2008hsf2.book..351W} that are not part of the YSOVAR standard set were selected based on an X-ray detection from observations other than \emph{Chandra} \citep[for the specific references see][]{2008hsf2.book..351W}.

\begin{figure}
\plotone{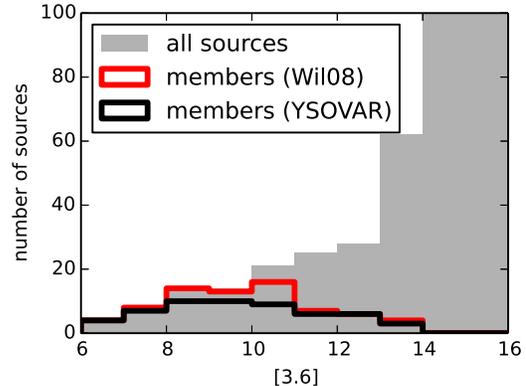}
\caption{Histrogram of observed mean IRAC~1 magnitudes. Grey is the entire set.
There are 126 and 261 sources in the range 14-15 mag and 15-16 mag, respectively, but the y-axis is scaled to smaller numbers for clarity. 
Shown in lines are the histograms for the subset of sources that are recognized as cluster members according to the YSOVAR membership criteria (black), 
or the work of \protect{\citet{2008hsf2.book..351W}} (red). \label{fig:hist36}}
\end{figure}
Figure~\ref{fig:hist36} shows a histogram of the observed mean magnitudes in $3.6\;\mu$m. Almost all bright sources are cluster members. The fraction of members drops below 11th~mag and no cluster member is found below 14th~mag. We expect background sources to be fainter than cluster members because of their larger distance and because they are seen through the cloud, but also fainter sources cannot be reliably classified. 
The thickness of the cloud is inhomogenous. The three primary fields cover roughly $A_V=20-35$~mag in the extinction maps from the c2d project \citep{2003PASP..115..965E}. This correspondes to $A_{[3.6]}=1.3-2.2$~mag \citep{1989ApJ...345..245C,2005ApJ...619..931I}. At the distance of L1688, a K5 star with little or no extinction has $m_{[3.6]} = 12.2$~mag according to the evolutionary tracks of \citet{2000A&A...358..593S} and the YSO colors from \citet{2013arXiv1307.2657P}. Thus, even late-K or early-M star members in the cloud are bright enough to be detected.

\subsection{Spectral slope}
\label{sect:spectralslope}

Just over 10\% of the 884 sources with lightcurves are classified by \citet{2009ApJS..184...18G} at all, because this classification scheme is conservative, representing a set of sources that can be dereddened and classified to a high degree of confidence. However, for many sources, we lack the required spectral coverage. Of those sources that can be classified in this way, 16 are class~I sources, 36 are class~II sources and 59 are class~III candidates, i.e., sources with a weak or absent IR excess in their SED (in this case the IR SED does not provide sufficient information to decide if a star is a YSO or a field star). Ten of those 59 have an X-ray counterpart in our \emph{Chandra} data.

In order to classify all sources, including those not classified in \citet{2009ApJS..184...18G}, paper~I defines a simpler approach, which uses the observed colors only. In this scheme, we fit the spectral slope $\alpha=\frac{d \log \lambda F_{\lambda}}{d \log \lambda}$, where $\lambda$ is the wavelength and $F_{\lambda}$ the flux density per unit wavelength interval at that wavelength. We make use of all measured flux densities (no upper limits) in the 2-24$\,\mu$m range, which corresponds to the range from the $K_s$ or $K$ filter, which we take from 2MASS and/or UKIDSS, respectively, to the MIPS 24$\,\mu$m channel. We use the flux densities for all \emph{Spitzer} bands from the observations presented in \citet{2009ApJS..184...18G} (and we include all sources extracted using this pipeline including those that are not published in \citet{2009ApJS..184...18G} because they are not classified as YSO in that work) or c2d and summarize the new lightcurves by calculating the mean and standard deviation for the new IRAC1 and IRAC2 data. Thus, sources detected both in the cryogenic mission and in the new dataset will have two datapoints for IRAC1 and IRAC2. 
If a source is detected in both 2MASS and UKIDSS we use both values for the fit, since they are independent measurements.
We perform a least-squares fit and call sources with $0.3 < \alpha$ class~I, $-0.3 < \alpha <0.3$ flat-spectrum (F), $-1.6<\alpha < -0.3$ class~II sources and $\alpha < -1.6$ class~III candidates. The classifications are given in Table~\ref{tab:tab2} in two columns. \texttt{s1\_SEDclass} presents the derived SED class using all available literature data as described above and the mean of the lightcurves from visibility window~1 for comparison with paper~I; \texttt{SEDclass} uses the same literature data but the mean flux density for the lightcurves calculated for all visibility windows.  In this scheme a significantly reddened main-sequence star, which has an intrinsic slope $\alpha<-1.6$, may appear as class~II object, so this observational classification cannot be translated directly into the evolutionary stage of an object. Of the remaining sources we find 110 class~I, 78 flat-spectrum and 455 class~II sources and 221 class~III candidates. We cannot classify 20 sources, because they are seen in one band only.

Not only the evolutionary status but also other properties of the individual source such as inclination or stellar mass influence the value of $\alpha$. Massive stars emit more energetic radiation and can thus change the structure of the accretion disk. However, L1688 is a region without massive YSOs \citep{2008hsf2.book..351W} and we define $\alpha$ using wavelengths longward of the $K$ band far from the peak of the stellar SED, so the shape of the photospheric SED has only negligible influence on the total SED. \citet{2006ApJS..167..256R} simulated different YSOs of low mass. Their results show that SEDs depend mostly on the evolutionary stage except for very extreme cases such as stars with an edge-on disk. Thus, $\alpha$ provides a good proxy for the evolutionary state of the YSOs in L1688.

Comparing with \citet{2009ApJS..184...18G}, the resulting classification is very similar, particularly for the class~I and flat-spectrum sources.
Most sources with a spectral slope $<0.3$ according to our slope-fitting are class~I sources in both classification schemes. With two exceptions our flat-spectrum sources are either class~I or II in \citet{2009ApJS..184...18G}. Of the 52 sources with a spectral slope between -1.6 and -0.3, 21 are also called class~II by \citet{2009ApJS..184...18G}, but 30 are class~III in that paper, indicating that a significant fraction of what we call class~II might indeed be reddened background stars. The general agreement between the more complex classification scheme and the observed spectral slope is also found for other star forming regions (paper~I).

\section{Mid-IR variability}
\label{sect:midirvariability}

We use three different methods to detect variability in all lightcurves, independent of their membership status or SED slope. Sources are considered variable, if they fullfill at least one of the following conditions: (i) A two-band lightcurve exisits and their Stetson index is larger than 0.9 (Sect.~\ref{sect:stetsonindex}); (ii) only a one-band lightcurve exisits and $\chi^2_\mathrm{red}>5$ (Sect.~\ref{sect:chi2}); (iii) the lightcurve is periodic (Sect.~\ref{sect:periodicity}).

\subsection{Stetson index}
\label{sect:stetsonindex}

\begin{figure}
\plotone{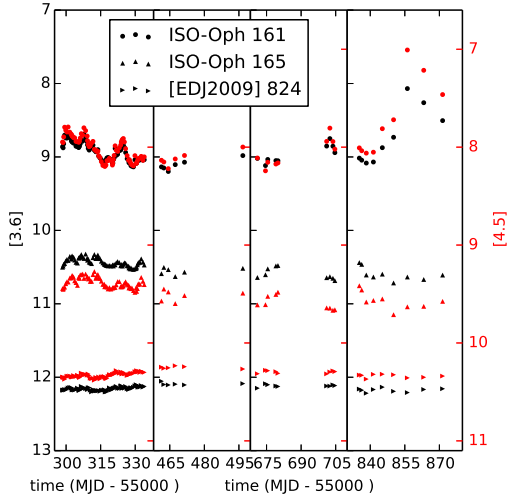}
\caption{Lightcurves for typical sources that are identified as variable using the Stetson index because the lightcurves in both bands are strongly correlated. 
The Stetson index is (from top to bottom): 13.1, 6.2, and 2.2.
Error bars are smaller than plot symbols. 
 \label{fig:stetsonlcs}}
\end{figure}

Sources in the target fields are observed in IRAC~1 and IRAC~2 almost simultaneously (within a few minutes). For those sources, we calculate the Stetson index $s$ with points weighted evenly \citep{1996PASP..108..851S}:
\begin{equation}
s = \frac{1}{\sqrt{N(N-1)}}\sum_{k=1}^{N}sig(P_k)\sqrt{|P_k|}
\label{eqn:stetson}
\end{equation}
where the sum is taken over all $N$ pairs of observations in IRAC1 and IRAC2 with observed magnitudes $a_k$ and $b_k$ and uncertainties $\sigma_{a_k}$ and $\sigma_{b_k}$; $sig$ denotes the sign of $P_k$. $P_k$ is the product of the normalized residuals in both bands:
\begin{equation}
P_k  = (\frac{a_k-\bar{a}}{\sigma_{a_k}}) \times (\frac{b_k-\bar{b}}{\sigma_{b_k}})
\end{equation}
Here $\bar{a}$ denotes the error-weighted mean of all IRAC~1 magnitudes and $\bar{b}$ the error-weighted mean of all IRAC~2 magnitudes. The Stetson index is very robust to observational errors since those are unlikely to affect both bands in the same way. Following paper~I, we define a source as variable if $s>0.9$. In paper~I, this is calculated over the first visibility window only; in this article we use the lightcurve from all four visibility windows. The Stetson index calculated for the first visibility window only can be found in Table~\ref{tab:tab2} in column \texttt{s1\_stetson\_36\_45} and for the entire lightcurve in column \texttt{stetson\_36\_45}. Figure \ref{fig:stetsonlcs} shows examples of lightcurves that are classified as variable according to the Stetson index. In the first visibility window, we find 34 sources to be variable, that is 18\% of all lightcurves where we can calculate the Stetson index. The number is similar (38 sources, 18\%) if we consider all observing visibility windows. This is not surprising, given that the Stetson index is designed to be a ``robust'' statistic that is little influenced by a small number of datapoints, and compared to the first visibility window, the other visibility windows only contribute a small number of datapoints. We caution that equation~\ref{eqn:stetson} will introduce a systematic bias in such a way that a source that is variable in only one band will not be recognized as variable. 

\subsection{$\chi^2$ test}
\label{sect:chi2test}

\begin{figure}
\plotone{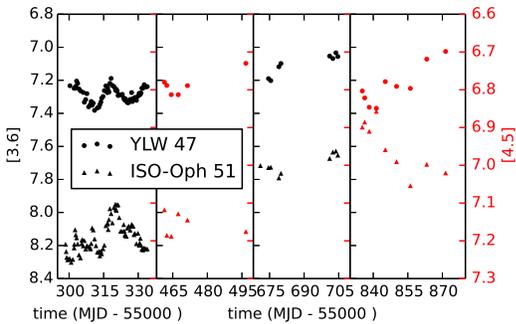}
\caption{Lightcurves for typical sources with useful observations in only one band per visibility window, where variability is detected according to the $\chi^2_{red}$ test. The reduced $\chi^2$ of the lightcurve is 50 and 240 for YLW~47 and ISO-Oph~51, respectively.
Both sources clearly show variability within a visibility window, as well as between visibility windows. Error bars are smaller than the plot symbols.\label{fig:chi2lc}}
\end{figure}
\label{sect:chi2}

For sources outside the primary target fields or for sources which are too bright or too faint in one IRAC channel, the Stetson index cannot be calculated. For those sources we rely on the reduced chi-squared value to detect variability:
\begin{equation}
\chi^2_{red} = \frac{1}{N-1}\sum_{k=1}^{N}\frac{(a_k-\bar{a})^2}{\sigma_k^2}
\end{equation}
Instrumental uncertainties lead to a non-Gaussian error distribution, so we use a conservative cut-off and mark sources as variable only if $\chi^2_{red} > 5$ (paper~I). Using this metric, we find 22 sources that exhibit variability in IRAC1 and 18 that exhibit variability in IRAC2. 
For comparison, we note that of the 38 sources classified as variable according to their Stetson index in the last section, 34 are also variable according to the $\chi^2$ test for their IRAC1 lightcurve and 37 according to their IRAC2 lightcurve.

The $\chi^2_{red}$ calculated for the first visibility window only can be found in Table~\ref{tab:tab2} in column \texttt{redchi2tomeans1\_36} and \texttt{redchi2tomeans1\_45} for IRAC~I and IRAC~2, respectively, and for the entire lightcurve in columns  \texttt{redchi2tomean\_36} and \texttt{redchi2tomean\_45}, again for IRAC~1 and IRAC~2, respectively. Examples are shown in Figure~\ref{fig:chi2lc}. If we fit a linear slope instead of comparing to the mean (i.e.\ a constant), the results are the same. 
%When fitting higher order polynomials, the number of sources with $\chi^2_{red} > 5$ goes down, e.g. for a quadratic polynomial only 29 and %18 sources have a $\chi^2_{red}> 5 $ in IRAC~1 and IRAC~2, respectively. 
Many sources have significantly larger values of $\chi^2_{red}$ for the whole observation series than within one visibility window, indicating a constant luminosity over 40~days, but a change over timescales between six months and two years.

\subsection{Periodicity}
\label{sect:periodicity}

\begin{deluxetable*}{ccccc}
\tablecaption{\label{tab:period} Periodic sources in the fast-cadence data.}
\tablehead{\colhead{name} & \colhead{channel\tablenotemark{a}} & \colhead{period [d]} & \colhead{FAP\tablenotemark{b}} & \colhead{SED class\tablenotemark{c}}}
\startdata
CFHTWIR-Oph 29 & [3.6]-[4.5] & 2.6 & 0.03 & F \\
ISO-Oph 138 & [3.6] & 3.3 & 0.00 & II \\
SSTYSV J162636.08-242404.2 & [3.6] & 4.1 & 0.00 & I \\
ISO-Oph 152 & [3.6] & 4.7 & 0.03 & II \\
{[GY92]} 30 & [3.6] & 14.5 & 0.00 & I \\
WL 11 & [4.5] & 3.0 & 0.01 & II \\
2MASS J16271881-2448523 & [4.5] & 6.4 & 0.00 & III \\
SSTYSV J162622.19-242352.2 & [4.5] & 6.1 & 0.00 & III \\
ISO-Oph 34 & [3.6] & 2.2 & 0.00 & F \\
ISO-Oph 33 & [4.5] & 2.4 & 0.02 & F \\
ROXN 41 & [3.6] & 6.5 & 0.00 & II \\
ISO-Oph 154 & [3.6] & 5.6 & 0.00 & II \\
ISO-Oph 124 & [3.6] & 3.4 & 0.00 & I \\
WL 13 & [3.6] & 10.7 & 0.00 & II \\
\enddata
\tablenotetext{a}{Periodicity might be present in more channels. In this table, the channel adopted as most reliable is given. See text for details.}\tablenotetext{b}{False alarm probability (see text).}\tablenotetext{c}{Observed SED class using the lightcurves from visibility window 1.}
\end{deluxetable*}

\begin{figure}
\plotone{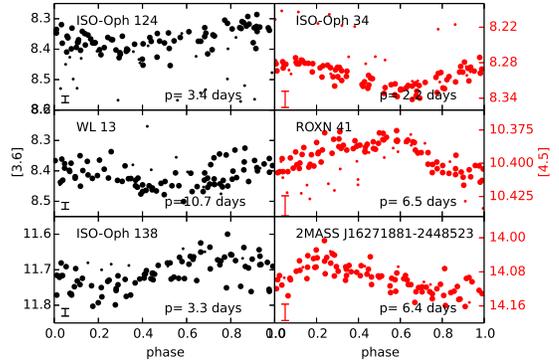}
\caption{Lightcurves for six periodic sources, folded by the period. Typical error bars are shown in each panel. 
Large dots are datapoints taken in the first visibility window. The different visibility windows are separated by about half a year. 
Given the limited accuracy of a period determined from the first visibility window only, 
phase-folded lightcurves from the other visibility windows (small dots) may appear with a phase-shift. 
Most lightcurves with a period in the first visibility window do not show a period in the later visibility windows. \label{fig:periodlc}}
\end{figure}

Lastly, we search for periodicity in the fast-cadence (visibility window 1) lightcurves using a Lomb-Scargle periodogram \citep{1976Ap&SS..39..447L,1982ApJ...263..835S}; again, see paper~I for details. We require a false alarm probability $<0.03$ and to further reduce the number of false positives, 
we additionally run a period detection of all lightcurves on the NASA
Exoplanet Archive Periodogram
Service\footnote{http://exoplanetarchive.ipac.caltech.edu/cgi-bin/Periodogram/nph-simpleupload}.
This service employs several algorithms because each algorithm is particularly suited to a different signal shape (see paper~I or the website for details on the other algorithms). We also calculate the autocorrelation function of each signal. We find that the Lomb-Scargle periodogram is advantageous for lightcurves with $<100$ points as is the case for most of our data. We require that at least one of three supplementary algorithms that we run retrieves a periodicity with a similar timescale (see paper~I for details).
We only search for periods between 0.1 and 14.5 days, so that at least three periods fit in each visibility window. Allowing longer periods leads to the detection of many long-term trends, where the data do not show that these trends are actually periodic. Finally, all algorithmically detected periodicity is vetted by eye. If periods are found in multiple bands, we report the period in IRAC~1, which is generally the most reliable due to the lower measurement uncertainties. If IRAC~1 does not reveal a periodicity, we report the value for the IRAC~2 lightcurve, and, as a last resort the period found in the [3.6]-[4.5] color to include objects where a periodic signal is overlayed by a long-term trend so that it is undetectable in each individual channel, but might be visible in the color.
The final list of adopted periods is given in Table~\ref{tab:period}. It contains sources from all SED classes.

We find 14 sources that show a periodic behavior in the first visibility window, but this periodicity is not stable over more than one visibility window. The datapoints taken in the first visibility window follow the folded period much better than the data from the later visibility windows, indicating that the period is not stable for longer than a few months. In ROXN~41, the datapoints of the later visibility windows, folded with the same period, seem to follow a different, yet clearly defined lightcurve with a similar period. 

Ten periodic sources have information in both IRAC bands. Eight of those are already classified as variable by the Stetson index. Three out of four sources with data in one band only are variable according to the $\chi^2_{red}$ test. The remaining two sources with information in both IRAC bands show periodicty, but fail the Stetson index test because a larger variability amplitude is required for a significant detection in Stetson index, which does not make any assumption about the form of the variability compared with the Lomb-Scargle periodigram, that only detects periodic signals. Equally, the remaining source with data in only one band fails the $\chi^2_{red}$ test, because the $\chi^2_{red}$ test also requires a larger variability amplitude for a significant detection than the Lomb-Scarge periodogram.

\begin{figure}
\plotone{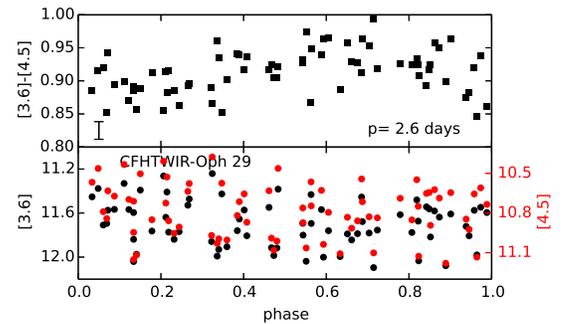}
\caption{Lightcurves for CFHTWIR-Oph~29 in each band separately (bottom) and in [3.6]-[4.5] color (top), folded by the period obtained from the [3.6]-[4.5] color. Typical error bars are shown in the upper panel; they are smaller than the plot symbol for the lower panel. 
\label{fig:periodcolorlc}}
\end{figure}

The number of datapoints in visibility windows 2, 3, and 4 is too low to search for periodicity in those visibility windows alone. Figure~\ref{fig:periodlc} shows the phase-folded lightcurve in one band for six periodic sources where the periodicity is significant in a single band. Figure~\ref{fig:periodcolorlc} shows the phased lightcurve for \object{CFHTWIR-Oph 29}, where the period is seen only in the color term [3.6]-[4.5]. Most of the periods found in Table~\ref{tab:period} are in the range 3-7 days. Only two sources have longer periods of 11 and 14 days. The largest amplitude is around 0.3~mag, and the smallest around 0.05~mag.

All periodic sources except  ROXN~41, SSTYSV J162636.08-242404.2, 2MASS J16271881-2448523, and SSTYSV J162622.19-242352.2 are cluster members according to our membership criteria. ROXN~41 and SSTYSV J162636.08-242404.2 have class I and II SEDs, respectively.
All but CFHTWIR-Oph~29 (see Appendix~\ref{sect:g09notwil}), SSTYSV J162636.08-242404.2 (see Appendix~\ref{sect:g09notwil}), and
2MASS J16271881-2448523 (no information beyond the $JHK$ magnitudes is available in the literature) are cluster members according to \citet{2008hsf2.book..351W}. 

Ten of the 56 cluster members are periodic; in total there are 14 periodic sources out of
60 sources
that are variable in the first visibility window. In comparison, \citet{2001AJ....121.3160C} find 18\% of all
variable stars to be periodic in $JHK_s$ monitoring of the Orion~A molecular cloud with a similar 
time coverage as we have for L1688. \citet{2013arXiv1309.5300P} find a third of all variable stars 
in L1688 to be periodic; due to their longer time baseline, \citet{2013arXiv1309.5300P} are sensitive
to different periods, but only 4 out of their 32 periods have values that are outside of the range we 
could detect -- see their table. We are sensitive to periods of up to 14.5~days here.

\subsection{Summary of variability detection}
\label{sect:summaryofvariabilitydetection}

The largest group of lightcurves in the sample of L1688 does not have any significant variability. The lower limit where variability is detected depends on the brightness of the object - fainter sources need a stronger relative variability due to the larger measurement uncertainties. Paper~I presents Monte-Carlo simulations to show that the Stetson test finds variability when the amplitude is a few times larger than the average uncertainty; the exact number depends on the signal shape. For example, the variability in a source that switches between two discrete levels will be detected when the amplitude is at least twice the uncertainty ($2\times0.015$~mag for a star with magnitude 14). If data from only one band are available, the step size must be more than four times the uncertainty to be found by the $\chi^2$-test ($4\times0.015$~mag for a star with magnitude 14.). Given the observing cadence, the Monte-Carlo simulations show that we are sensitive to periods between 0.1 and 14.5 days (paper~I).

In summary, we call a source variable if it is either periodic with a low false alarm probability (14 sources)
or fulfills one of the following conditions: 
Sources with simultaneous data in two bands need to have a Stetson index $>0.9$ 
(34 sources in the YSOVAR standard set fast cadence data and 38 sources in total)
and sources without simultaneous data need to have $\chi^2_{red} > 5$ 
(22 sources in the YSOVAR standard set fast cadence data and 29 sources in total).

\section{Results and Discussion}
\label{sect:resultsanddiscussion}

We consider the lightcurves for all 882 distinct sources with at least five datapoints in at least one IRAC band. 
Of those lightcurves, 70 are classified as variable; 56 sources are cluster members according to the YSOVAR criteria (44 of them are variable) and
73 sources are cluster members according to \citet{2008hsf2.book..351W} (47 of them are variable). Both membership samples have considerable overlap.

In the following subsections, we compare properties of the lightcurves between different SED classes. 
In most cases, the lightcurves of IRAC1 (3.6$\,\mu$m) and IRAC2 ($4.5\,\mu$m) have very similar properties.

\subsection{Evolutionary trends of the variability}
\label{sect:evolutionarytrendsofthevariability}

\begin{figure}
\plotone{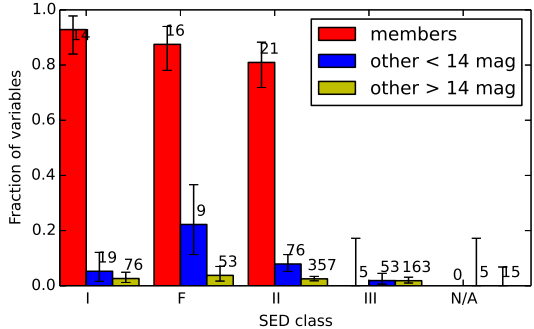}
\caption{Fraction of variable sources in each SED class, split between cluster members and non-members. 
The numbers on the bar chart give the total number of sources in that category; the height of the bar is the fraction of variables.
Given a finite sample size, the observed fraction of variables may differ from the true fraction, which is shown by the error bars (see text for details).
If no variable source is found, then no barchart is drawn and only the error bar and the number are visible, e.g.\ for objects with an unclassifiable SED (N/A) due to missing data.
Source are part of the bright sample, if $[3.6] < 14$ or $[4.5] < 14$.
\label{fig:probvarSED}}
\end{figure}
\begin{figure}
\plotone{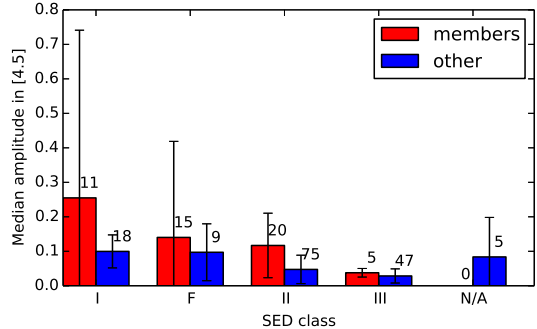}
\caption{Median amplitude in magnitudes of $4.5\;\mu$m lightcurves in each SED class; the value for the amplitude is the difference between the 10\% and the 90\% quantile for each lightcurve.
Only sources with a mean magnitude $<14$ are shown here, where the uncertainty on each datapoint is small compared to the intrinsic variation.
The numbers on the bar chart give the total number of sources in that category.
The error bars shown are the standard deviation of the amplitude for all sources in that class.
\label{fig:meanampSED}}
\end{figure}
Figure~\ref{fig:probvarSED} contains the variability fraction sorted by SED class (see section~\ref{sect:spectralslope}). Within each class, the population is sub-divided into previously identified members, bright ($<14$~mag in $3.6\;\mu$m or $4.5\;\mu$m) stars not previously identified as members, and faint ($>14$~mag) stars not previously identified as members.
Sources are considered variable if they have a low false alarm probability for periodicity or a Stetson index $>0.9$ (if simultaneous data in IRAC~1 and IRAC~2 exist) or $\chi^2_{red}>5$ (if no simultaneous data exist). Note that the classes shown here are based on the observed SED and thus background non-member stars seen through the cloud might appear red (like a class II source) even if their intrinsic SED slope is compatible with SED class~III. No star classified as a member is fainter than 14~mag. 

The observed fraction of variable stars as a function of SED class is the best guess for the probability to find variability for a star of a given SED class. We calculate the uncertainty for the probability based on the observed number of variable stars $v$ in a bin with $N$ total sources as follows: If the true probability for a source in the bin to be variable is $p$, then the probability to observe $v$ variables in $N$ stars is given by $p_{obs}$, which follows a binomial distribution:
\begin{equation}
p_{obs}(v,N,p) = {N \choose v} p^v (1-p)^{N-v}
\end{equation}
The peak of this distribution is $p_{obs}(p) = \frac{v}{N}$. We then calculate the boundaries of the confidence interval $p_1 .. p_2$ for $p$ such that $p_{obs}(p_1) = p_{obs}(p_2)$ and the area under the curve includes 68\% of the probability. Those uncertainty ranges are shown in Figure~\ref{fig:probvarSED}.

Almost all known cluster members are variable; the fraction of variables is decreasing slightly from class~I to II. There are five class~III members and none of them shows variability. This is compatible with the variability fraction observed for class~III non-members. By definition, sources with a class~III SED do not have a large IR excess over a stellar photosphere, thus they are not expected to harbor a substantial disk or show disk-related variability. However, even field stars can show significant variability, at least in the near-IR. \citet{2013AJ....145..113W} find 1.6\% of all field stars to be variable in $JHK$. This includes eclipsing systems, stars with unusually large photospheric spots and other unidentified variables. All effects that influence the $JHK$ lightcurve are likely also visible in the mid-IR. Our fraction of variable objects with a class~III SED is similar (1.8\%) to what Wolk et al.\ found. 
In the optical and near-IR, the fraction of variable field stars is lower than that value \citep{2009A&A...503..651P,2012A&A...537A.116P,2013AJ....145..113W}.

\emph{Spitzer} observations of other young clusters also find that younger stars are more variable, but with a lower variability fraction. \citet{2011ApJ...733...50M} find only 70\% of all stars with disks in their ONC sample to have detectable variability. \citet{2013AJ....145...66F} find 60\% in IC~348 and \citet{cody2014} find $>90$\% of all members in NGC~2264 to be variable in the IRAC bands. All three clusters are located at larger distances and thus observations are less sensitive to small variations, the observations of \citet{cody2014} are more densely sampled and thus provide a better signal. In L1688, a K5 star with little extinction will have a magnitude of 12.2 in $3.6\,\mu$m using the evolutionary tracks of \citet{2000A&A...358..593S} and the YSO colors from \citet{2013arXiv1307.2657P}. In this magnitude range, variability down to 0.05~mag can be detected (paper~I). In contrast, for the same star in IC~348, the variability has to be about twice as large to be detected and three times as large to the detected in the ONC. 
Of course, even our observations miss the faint end of the distribution. Thus, it seems likely that essentially all YSOs show substantial mid-IR variability on timescales of days to weeks.

Sources with a class~III SED are not contaminated by disk-bearing stars, since reddening by the cloud can only increase the spectral slope $\alpha$ but never hide an existing IR excess. The larger fraction of non-member variable objects in the other classes in Figure~\ref{fig:probvarSED} thus shows that the membership lists are incomplete. One caveat here is that the class~III sources in the sample of members are selected in a different way than the other classes. If true class~III cluster members with X-ray emission (that are included in our member sample) are systematically less variable than class~III members where we do not detect X-ray emission (that are therefore not included in the member sample), that would also lead to a lower observed variability fraction in class~III member sources. However, given the size of the observed effect in variability such a bias seems to be unlikely to be the sole reason for the observed distribution.

The observed probability that a source is variable for all classes of bright non-members (blue bars) is consistent with around 5\%. The fact that this barely changes with the SED class, quite unlike the distribution for the cluster members, indicates that unidentified cluster members cannot make up a large fraction of the non-member sample since they would bias the observed variability fraction to higher values for the earlier classes. The vastly different number of sources in the different bins makes it highly unlikely that each bin is contaminated by the same fraction of class~I-II cluster members.
The sample of faint stars has almost the same fraction of variable sources in every SED class. The low fraction can be explained by two effects. First, variability of faint sources cannot be reliably detected unless the amplitude is exceptionally large. Second, as most cluster members are brighter than 14~mag, the faint sample contains fewer unrecognized cluster members than the sample of brighter sources.

Figure~\ref{fig:meanampSED} shows that there is a wide distribution of amplitudes in YSOs of class I, F and II. The error bars in this figure represent the standard deviation of the mean amplitudes within one class. Amplitude and standard deviation in the figure are measured in magnitudes; a median amplitude of 0.1~mag means that the flux in $4.5\;\mu$m varies by 10\% for a typical source.
For members, the variability amplitude is larger in class~I sources than in flat-spectrum and class~II sources, where the amplitude is still larger than in class~III sources. Also, the spread of the amplitudes in class I and F is much larger than for class II and III. For non-members, the observed amplitudes seem to follow a similar, but less pronounced trend. This suggests that the non-member category still includes not only reddened background objects with an intrinsic class~III SED, but also some unidentified cluster members.
Indeed, we propose that all variable class~I, F and II sources are members and Table~\ref{tab:newmembers} lists those new members that are neither contained in our YSOVAR standard membership set (see Table~\ref{tab:tab2}) nor in \citet{2008hsf2.book..351W}. However, for consistency, we do not modify our sample of members at this point and continue to treat the sources in the table as non-members for the remainder of the analysis. Two of the objects in the table, \object{CFHTWIR-Oph 21} and \object{CFHTWIR-Oph 74} were suggested as a $\rho$~Oph substellar candidate members by \citet{2010A&A...515A..75A}, the remaining objects have not been studied before.

In summary, the probability for any given source to be observed as a mid-IR variable decreases little between class~I and II sources, but the mean amplitude as well as the differences within a class are much larger for sources which have more circumstellar material.

\begin{deluxetable}{cccc}
\tablecaption{New members for L1688 in addition to the YSOVAR standard member set and \protect{\citet{2008hsf2.book..351W}} \label{tab:newmembers}}
\tablehead{\colhead{RA} & \colhead{DEC} & \colhead{name} & \colhead{SED}}
\startdata
246.88244 & -24.79978 & SSTYSV J162731.78-244759.2 & II \\
246.64060 & -24.31667 & SSTYSV J162633.74-241900.0 & II \\
246.84236 & -24.78363 & SSTYSV J162722.16-244701.0 & II \\
246.59001 & -24.49297 & SSTYSV J162621.60-242934.7 & II \\
246.65036 & -24.40117 & SSTYSV J162636.08-242404.2 & I \\
246.84092 & -24.31180 & SSTYSV J162721.82-241842.4 & II \\
246.84135 & -24.78713 & SSTYSV J162721.92-244713.6 & II \\
246.84336 & -24.64383 & CFHTWIR-Oph 74 & II \\
246.90225 & -24.64854 & SSTYSV J162736.53-243854.7 & II \\
246.57278 & -24.38732 & SSTYSV J162617.46-242314.3 & II \\
246.67446 & -24.25616 & SSTYSV J162641.86-241522.1 & II \\
246.91134 & -24.55516 & SSTYSV J162738.72-243318.5 & II \\
246.60121 & -24.26383 & CFHTWIR-Oph 21 & F \\
246.88281 & -24.34855 & SSTYSV J162731.87-242054.7 & F \\
246.86474 & -24.43647 & SSTYSV J162727.53-242611.2 & I \\
246.59027 & -24.30560 & SSTYSV J162621.66-241820.1 & F \\
246.86794 & -24.67487 & SSTYSV J162728.30-244029.5 & I \\
\enddata
\end{deluxetable}

\subsection{Timescale of variability}
\label{sect:timescaleofvariability}

Next, we will look at the timescales of the lightcurves. The list of periodic sources contains objects of all evolutionary stages, but is too small to recognize any trends in the period with evolutionary stage. We calculate the discrete auto-correlation function $ACF(\tau)$ for each fast-cadence (visibility window 1) lightcurve to characterise the time-scale in all sources, not just the periodic ones. One complication here is that our data are unevenly sampled. In order to calculate the $ACF$ we linearly interpolate the lightcurve on a grid with time steps of 0.1~days. This process can change the properties of the lightcurves on short timescales. However, for most lightcurves the relevant timescales are longer than the distance between two observations (0.1 to 0.8 days). We use the following definition of the $ACF$
\begin{equation}
ACF(\tau) = \frac{1}{N-\tau}\sum_{k=1}^{N-\tau}\frac{(a_k-\bar{a})(a_{k+\tau}-\bar{a})}{\sigma_a^2}
\end{equation}
where $N$ is the total number of points in the lightcurve $a_1, a_2, ... , a_N$ with a mean of $\bar{a}$ and a standard deviation of $\sigma_a$. In the discrete $ACF$ $\tau$ is the number of timesteps.
\begin{figure}
\plotone{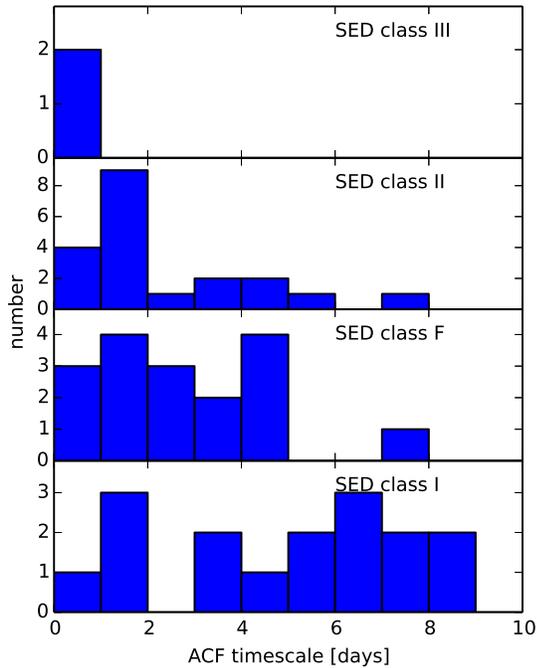}
\caption{Histogram of the timescale $\tau$ in the auto-correlation function (ACF) for variable sources in different SED classes, such that $\tau$ is the smallest tine scale with $ACF(\tau)<0.25$.
\label{fig:ACF}}
\end{figure}
By definition $ACF(\tau=0)=1$ and the $ACF$ decays for longer time-lags $\tau$. We take the first value of $\tau$ with $ACF(\tau)<0.5$ as the characteristic time-scale for a lightcurve. A more common definition would be to use the position of the first local maximum in the $ACF$, but due to the low number of datapoints in our lightcurves, the noise in the $ACF$ is large and this value is often not well defined. While our definition might not give us the timescale of the physical processes at work, the $ACF$ timescale still provides relative comparisons. We find that the average $ACF$ timescale decreases from 4~days for class I sources to one day for class~III sources (Figure~\ref{fig:ACF}). A more detailed analysis of stochastic and quasi-periodic properties of the lightcurves is beyond the scope of the current paper \citep[for a discussion of these properties in $JHK_s$ lightcurves, see][]{2013arXiv1309.5300P}.

\subsection{Morphological types of variability}
\label{sect:morphologicaltypesofvariability}

The majority of the stars included in our data show no
significant variability, and we exclude them from further
discussion. The remaining stars are definitely variable,
with light curve shapes that display a variety of 
morphologies.  Visual inspection, and some quantitative
analysis, allows us to group these stars into sets with
similar light curve morphologies - which can be the first
step in attaching physical mechanisms as the cause of the
observed variability.  Based on a near-IR ($JHK$) monitoring
campaign of similar cadence and duration to ours, \citet{2013ApJ...773..145W} identified four light curve morphological classes:
(a) periodic;
(b) quasi-periodic - which they defined as stars with cyclic
  brightening and fading, but where the frequency and amplitude
  of the variations varied from cycle to cycle;
(c) long-duration - stars with relatively long term monotonic
    changes in brightness over weeks or months, with eventual
    changes in sign of the variability;
(d) stochastic - which indicated all other variability types,
   where no obvious pattern to the variability was present.

In another recent paper, \citet{cody2014} analysed a $30+$ day
monitoring campaign for the star-forming region NGC 2264, using
optical data from CoRoT and IR data from IRAC, to assign
light curve morphology classes to their young star set.  Their
proposed light curve morphologies were:
(i) periodic - stars whose light curves show periodic waveforms
   whose amplitude and shape are unchanging or only change in 
   minor ways over the 30 day observing window.  These light 
   curves were ascribed generally to cold spots on the stellar
   photosphere;
(ii) dippers - stars showing a well-defined maximum brightness,
  upon which are superposed flux dips of variable shape and
  amplitude.  These are sub-divided into stars where the flux
  dips occur at an approximately constant period - designated
  as quasi-periodic systems - and stars with no obvious
  periodicity to the dips - designated aperiodic.  In previous
  literature, these two sub-classes, or portions of them, have been 
  referred to as AA Tau systems and UX Ori systems, respectively;
(iii) short duration bursters - stars with relatively well-defined
  minimum light curve levels, superposed on which are brief (hours
  to day) flux increases, attributed to accretion bursts.  See
  \citet{2014arXiv1401.6600S} for further discussion of this set;
(iv) quasi-periodic variables - stars lacking a well-defined 
  maximum brightness but showing periodic variability whose waveform
  changes shape and amplitude from cycle to cycle;
(v) stochastic - stars with prominent luminosity changes on a 
  variety of timescales, with no preference for ``up'' or ``down''.
(vi) long timescale variability - stars with slow (weeks to
  months) changes in brightness.

While there are some clear similarities in the two schemes, their
usage of the terms quasi-periodic and stochastic are not the same,
and light curves described as belonging to those classes in one scheme
would not necessarily be so classified in the other scheme.  It
will be important to resolve these nomenclature issues in the future,
perhaps in the same manner as was done for sorting out similar
issues on classifying pre-main sequence disk SED morphologies
(Evans et al. 2009).  In the meantime, we must choose which scheme
to adopt and be clear that is what we have done.   For this paper,
we adopt the \citet{2013ApJ...773..145W} scheme. 

\begin{figure*}
\plotone{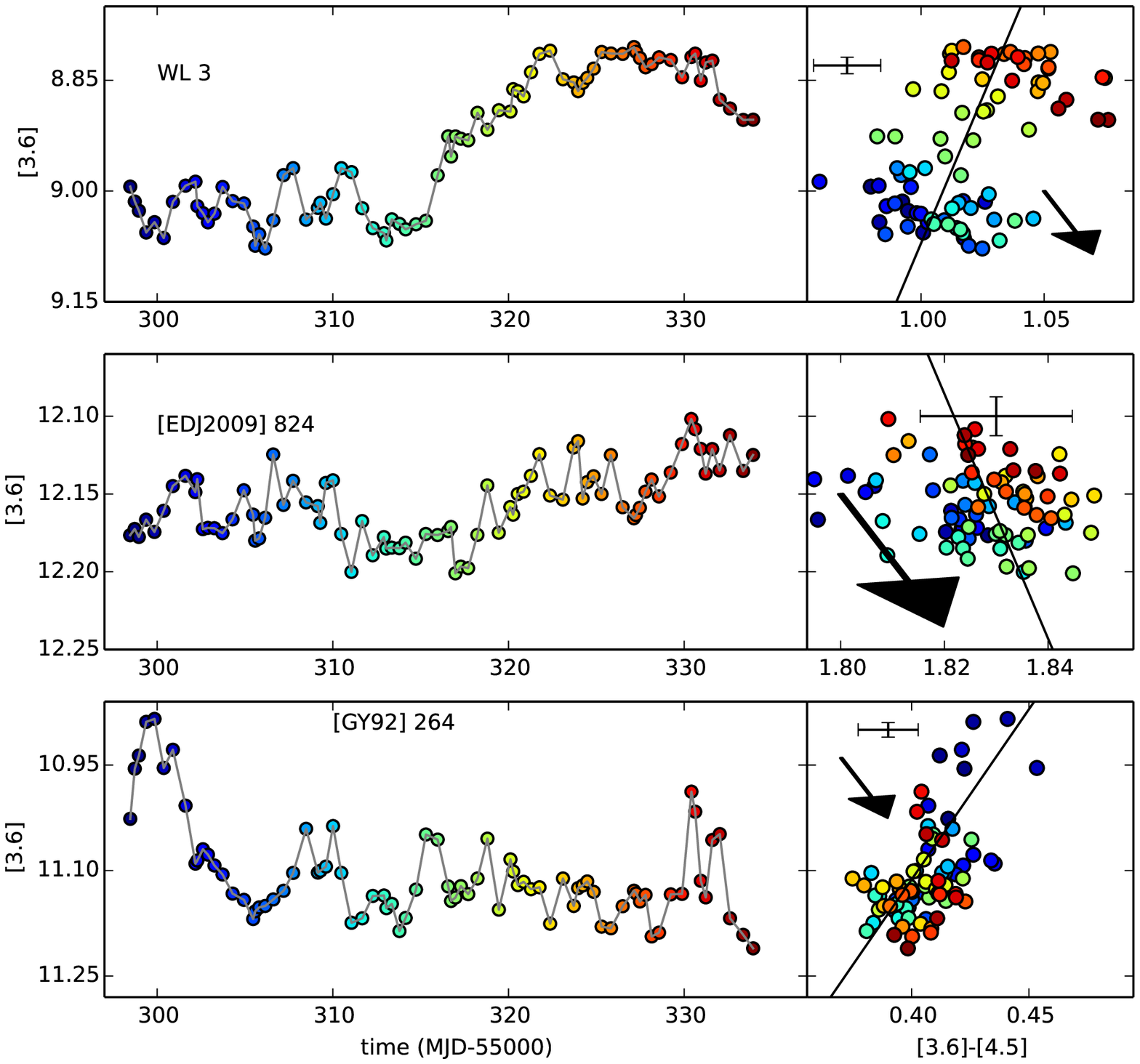}
\caption{Examples for lightcurves with quasi-periodicity. Lightcurve and color-magnitude diagram (CMD) in the first visibility window for sources with two-band coverage. 
The color of the plot symbols in the CMD shows which part of the lightcurve is responsible for each point in the CMD. 
Typical error bars are shown in each CMD. The line in the CMD marks the best fit through all
datapoints. The arrow indicates a reddening of $A_K = 0.1$~mag \citep{2005ApJ...619..931I}. 
Note that the x and y axes have different scalings which make the slope appear less steep than it is.
\label{fig:morelcs}}
\end{figure*}

Periodic lightcurves are described in section~\ref{sect:midirvariability} and examples can be seen in Figure~\ref{fig:periodlc}. Examples of quasi-periodic lightcurves are shown in Figure~\ref{fig:morelcs}. \object{WL 3} shows variability with a timescale $\sim3$~days in the first half of the first visibility window, but around MJD 55316, the flux increases significantly and this rise masks out any periodicity. The dip around MJD~55325 again has a similar duration as those observed in the beginning of the visibility window. The other two sources shown might be similar, but the signal-to-noise is not as good. 

\begin{figure}
\plotone{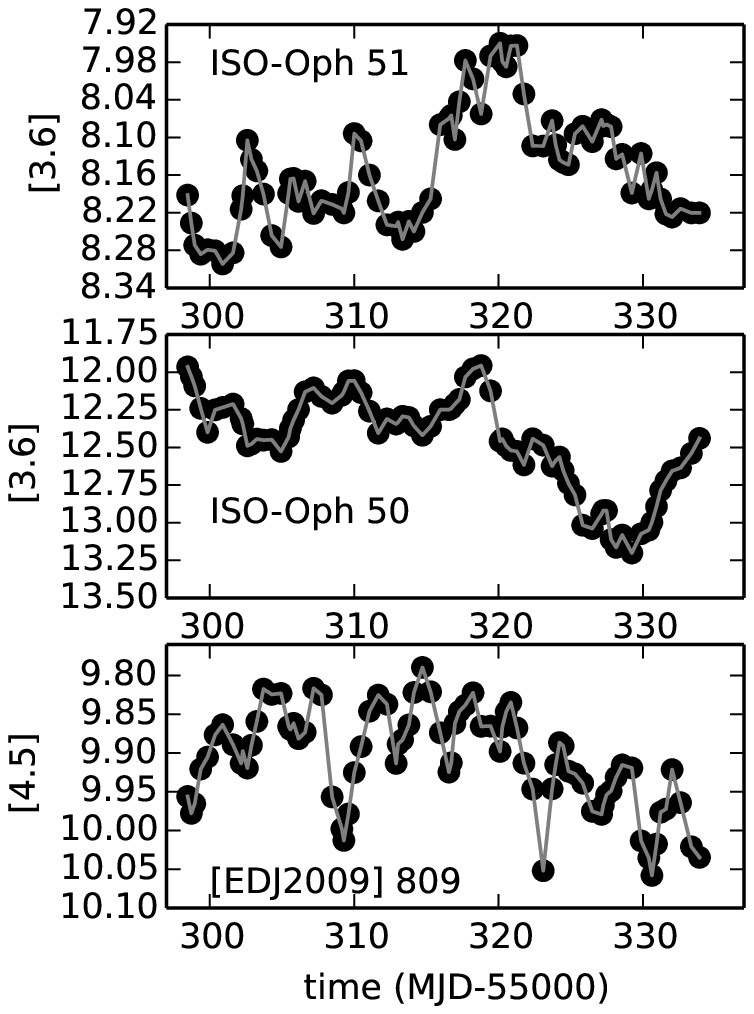}
\caption{Examples for aperiodic lightcurves in one band. Errorbars are smaller than or comparable to the plot symbols.
\label{fig:morelcs2}}
\end{figure}
Other sources are variable, but no preferred timescale for the variability is discernible. \object{ISO-Oph 50} and 51 (Figure~\ref{fig:morelcs2}) are examples of this aperiodic behavior. In yet other cases, the timescale of the variability is so long that we cannot decide if a feature is a singular event or part of a recurring pattern. Good examples of this are WL~4 and \object{[EDJ2009] 892}, which are shown in Figure~\ref{fig:cmd}.

In some cases, we see short duration features in the lightcurve in addition to a longer trend. This can either be a brightening (e.g.\ MJD 55303 and 55311 in \object{ISO-Oph 51}, Figure~\ref{fig:morelcs2}) or short dips in the lightcurve that last between one and ten days (\object{[EDJ2009] 809} in the same figure).

Like \citet{2013ApJ...773..145W}, we find that even strongly periodic sources are not perfect clocks in the mid-IR in that the amplitude can vary from cycle to cycle.  There might also be phase shifts from visibility window to visibility window. In a very similar study to the one we present here, \citet{2013AJ....145...66F} observed IC 348 over a 40 day window with IRAC1 and 2 and find 25\% of the variable stars are likely periodic.  Here we find a similar division; 10 of 52 members are strongly periodic.

\subsection{Long-term variability}
\label{sect:longtermvariability}

\begin{figure*}
\plotone{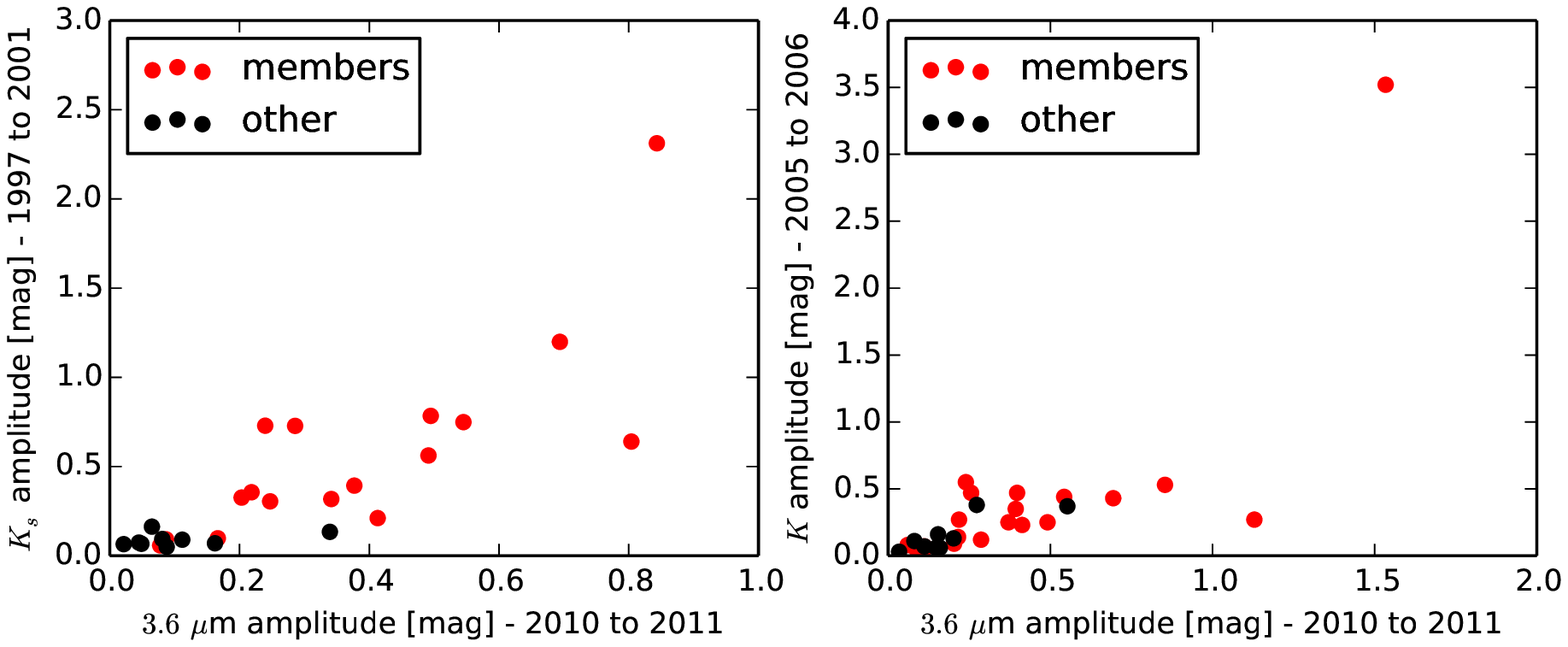}
\caption{Observed [3.6] amplitude in our monitoring compared with the observed $K_s$ or $K$ amplitude from the literature. \emph{left:} 1997-2001 \protect{\citep{2013arXiv1309.5300P}} and \emph{right:} 2005-2006 \protect{\citep{2008A&A...485..155A}}.
\label{fig:amplitudeKvs36}}
\end{figure*}
We present the first mid-IR monitoring of L1688, but the field has been monitored before in the near-IR. Figure~\ref{fig:amplitudeKvs36} compares the $K_s$ band amplitude found in 2MASS data taken between 1997 and 2001 \citep{2013arXiv1309.5300P} and the $K$ band amplitude observed at UKIRT between 2005 and 2006 \citep{2008A&A...485..155A} with the amplitude of our $3.6\;\mu$m lightcurves. Given the different bands, it is not surprising that the absolute value of the amplitude differs, but there is a good correlation such that the sources with the largest $K$ or $K_s$ band amplitudes in earlier observations also have the largest amplitudes in the IRAC bands five to ten years later. In most objects, color changes are small or happen in parallel with luminosity changes (section~\ref{sect:colorchangesandreddening}), so the $K$ or $K_s$ band and the [3.6] variability should be strongly correlated. Figure~\ref{fig:amplitudeKvs36} shows that the amplitude of the variability is relatively stable over at least one decade. \citet{2013arXiv1309.5300P} used a larger number of observations than \citet{2008A&A...485..155A} and observe a larger spread between a typical YSO's brightest and faintest $K$-band magnitude, indicating that the longest timescale of variability is longer than the time span of the \citet{2008A&A...485..155A} observations.

\citet{2013AJ....145...66F} noted that roughly 6\% of the stars in IC~348 had data which indicated significant changes in the source's mid-IR flux over the three year interval since the last IRAC observation.  \citet{2013ApJ...773..145W} also displayed numerous examples of stars which showed continuous change over the course of at least one observing visibility window. 

All this shows that the amplitude of variability of a given source may evolve over very long timescales, but is consistent over at least one decade. YSOs apparently do not switch between highly variable and much less variable states.

\subsection{Color changes and reddening}
\label{sect:colorchangesandreddening}

\begin{figure*}
\plotone{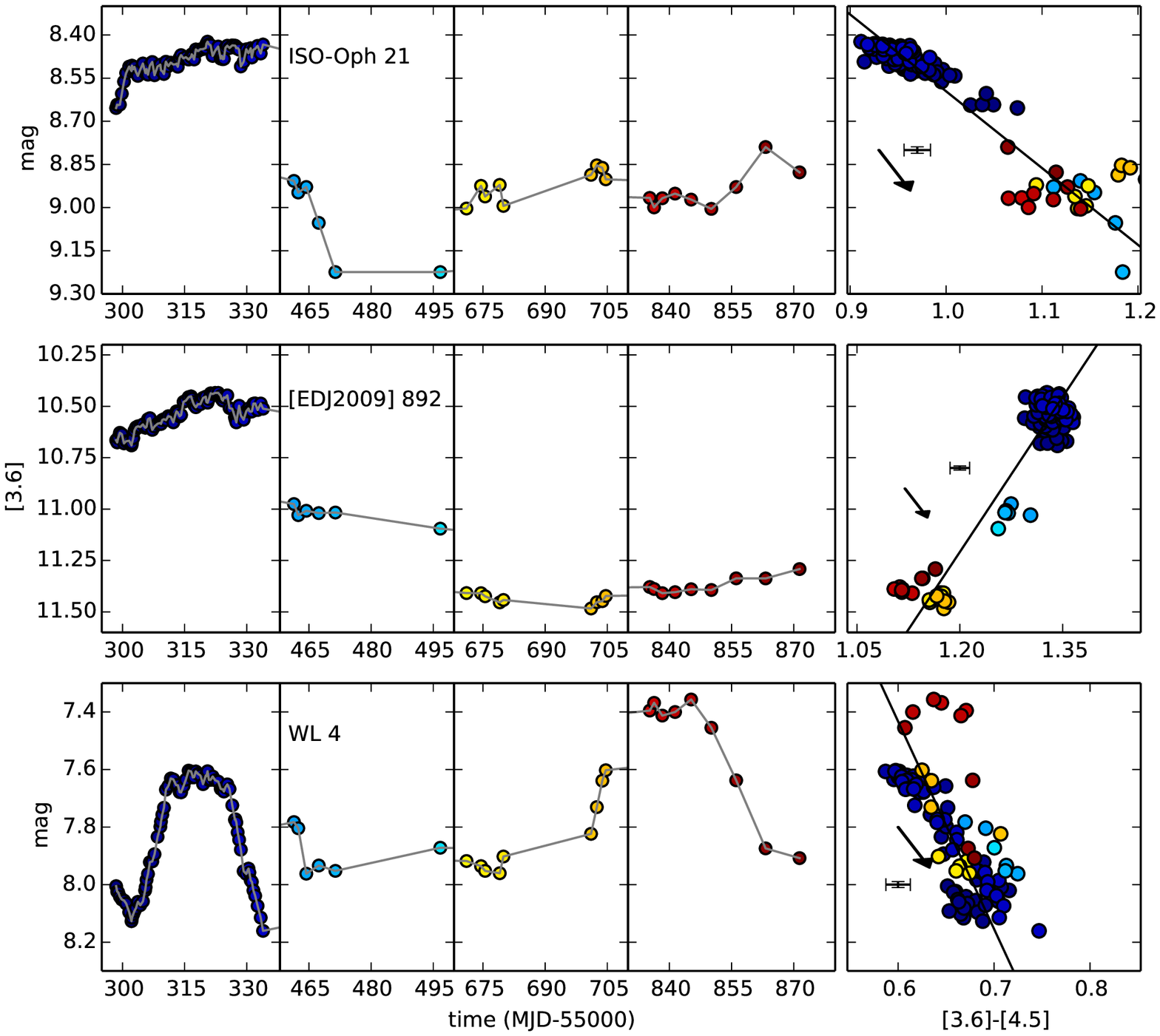}
\caption{Lightcurves and CMDs with long timescale variability. In contrast to Figure~\ref{fig:morelcs}, here lightcurves are shown for all observing visibility windows. Symbols in different colors are datapoints from different observing visibility windows. 
Typical error bars are shown in each CMD. The line in the CMD marks the best fit through all datapoints. 
Note that the x and y axes have different scalings which make the slope appear less steep than it is.
The arrow indicates a reddening of $A_K = 0.2$~mag \citep{2005ApJ...619..931I}.
Color variability can happen both between visibility windows (first and second row) or within a visibility window (bottom). 
Some sources show a slope in the CMD that is very close to the standard reddening (WL~4 and ISO-Oph~21); others appear bluer when they are dim ([EDJ2009]~892).
\label{fig:cmd}}
\end{figure*}
\begin{figure}
\includegraphics[width=0.4\textwidth]{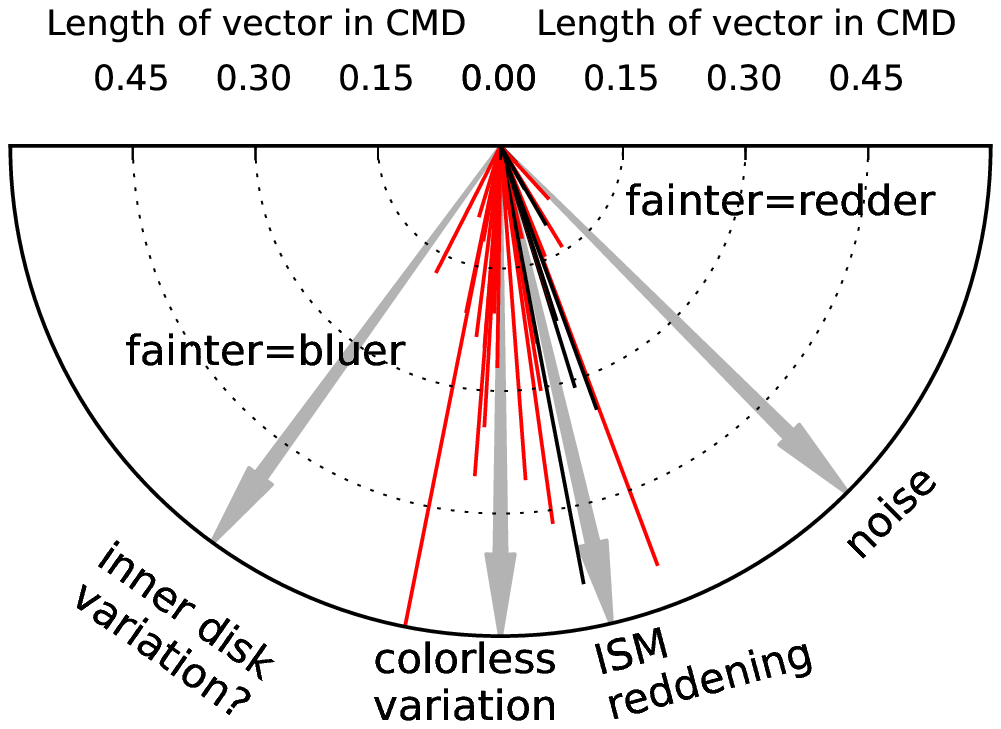}
\includegraphics[width=0.4\textwidth]{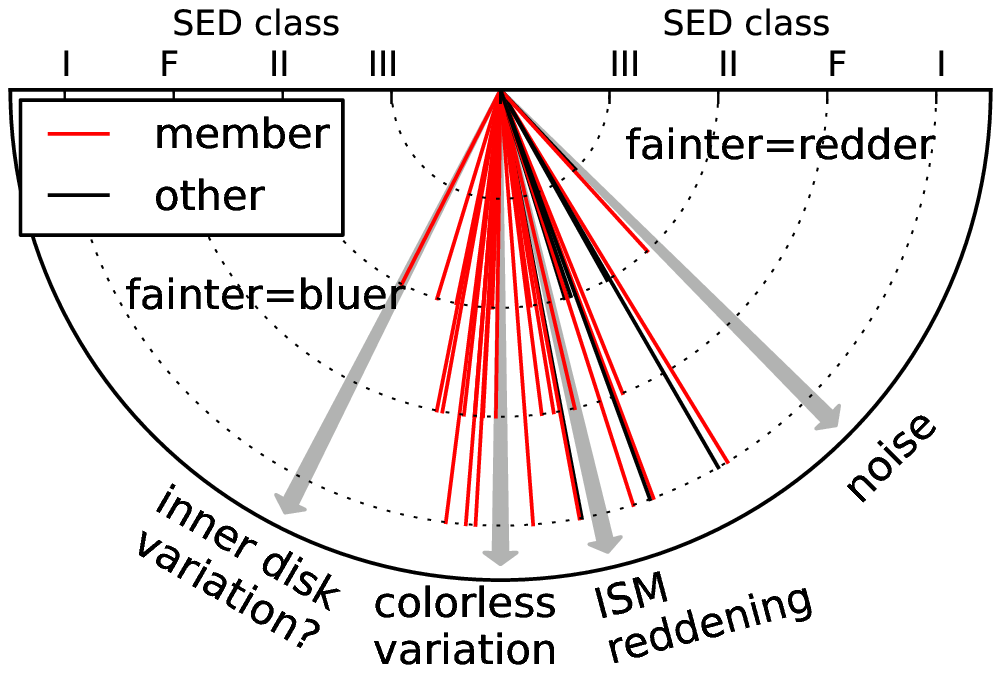}
\caption{Angle of the variability in the CMD for all sources with more than 10 simultaneous datapoints, where the uncertainty of the angle of a line in the CMD is below 6\degr.
A gray arrow marked ``ISM reddening'' indicates the reddening slope of \citet{2005ApJ...619..931I}.
\emph{top}:
The length of each line in this panel shows the amplitude of variability in a CMD.
In the CMD of each source all datapoints are projected on the best-fit line. The 10\% quantile and the 90\% on this line are calculated and the panel shows the distance between those two quantiles in mag. 
For CMDs with reddening this can be interpreted as the $A_{3.6}$ of time-variable extinction. 
The largest $A_{3.6}$ observed is 1, but for clearity only the region to $A_{3.6}=0.6$~mag is shown.
\emph{bottom}: 
In this panel the length of each line gives the SED class of that object. 
Only cluster members have slopes where the source bluens as it fades.
\label{fig:cmdslope}}
\end{figure}
For sources in the primary target fields, the observations in $3.6\,\mu$m and $4.5\,\mu$m are separated by only a few minutes. For those sources, we compare the color and the magnitude in the $3.6\,\mu$m band in a color-magnitude diagram (CMD) (Figures~\ref{fig:morelcs} and \ref{fig:cmd}). In some cases, the color is fairly stable within one visibility window, but changes between visibility windows; in others the timescale of variability is much shorter and color changes are seen within one visibility window as well. For non-variable sources, the CMD forms a point cloud with the size set by the photometric uncertainties, but for variable sources, the shape of the CMD can reveal the physical cause of the variability -- for example, if a disk warp or accretion funnel passes in front of a YSO, we expect its color to become redder as it becomes fainter. If the absorber has the same gas and dust properties as the inter-stellar medium (ISM) and absorbs star and inner disk at the same time, then the datapoints in the CMD follow a line with the slope of the interstellar reddening law \citep[e.g.][]{2005ApJ...619..931I}.

In each CMD, the observational data (I1, I1-I2) are fit to a line segment using an orthogonal distance regression method that takes the errors in both the x and y directions into account. We calculate the length in magnitudes of the line segment excluding the 10\% of the data that are outliers on either side.
The CMD slopes of the 37 sources with a well-defined slope are shown in Figure~\ref{fig:cmdslope}. The x and the y axis in the CMD are correlated, because they both depend on the $3.6\,\mu$m magnitude so sources which are noise-dominated would show a slope of $45^{\circ}$. No source is seen in this region.

From a total of 42 variable sources with CMDs with at least ten datapoints, 37 can be fitted with a formal uncertainty on the best fit slope below $6^{\circ}$. In four of the five sources with a statistical error on the slope in the CMD $>6^{\circ}$, the color variability is dominated by observational uncertainties; the remaining source ISO-Oph~140 is discussed below.

As can be seen in Figure~\ref{fig:cmd}, even in those sources with a well defined slope in the CMD, the scatter around the best linear fit is larger than the measurement uncertainties; individual visibility windows are systematically above or below the fitted line. 

The slopes shown in Figure~\ref{fig:cmdslope} can be separated (somewhat arbitrarily) in two groups. One group becomes redder when the sources are fainter with slopes comparable to the ISM reddening. This group contains almost all variables that are not classified as cluster members in section~\ref{sect:l1688membership}. Most sources in this group have class~I or flat-spectrum SEDs (bottom panel).

Sources that bluen (we use the term ``bluen'' as a verb to mean that a source comes bluer similar to the common expressions ``redden'' or ``reddening'') when they are fainter are mostly class~II and flat spectrum sources. Figure~\ref{fig:morelcs} shows the lightcurves and CMDs for two of the bluening sources, \object{WL~3} and \object{[GY92] 264}.

The top panel in Figure~\ref{fig:cmdslope} quantifies the magnitude of the reddening or bluening observed in each source.  For a source where the slope in the CMD is compatible with ISM reddening, this can be interpreted as the $A_{3.6}$ of the intervening material. A range of values is observed, but in most sources it is $<0.3$~mag. No significant difference in vector length is seen between sources that bluen or redden when they dim.
We discuss the physical mechanisms that could cause these slopes in section~\ref{sect:scenariosforcolorchanges}.

Comparing the amplitudes of all sources in Figure~\ref{fig:cmdslope} that either redden or bluen over the full time span of the observations, we do not find significant differences in value or distribution of the variability amplitudes.

Similar to our results, \citet{2008A&A...485..155A} find reddening and bluening slopes in the CMD in their $HK$ monitoring, but compared to Figure~\ref{fig:cmdslope} they see a higher fraction of stars that bluen as they dim (33/49). Only six sources have well defined slopes in both their study and ours. In the data of \citet{2008A&A...485..155A}, all of them belong to the group that bluen with lower fluxes. In five cases, our data agree. In contrast, ROXN~44 reddens in our observations, so its behavior changed over the time interval of five years, or it is behaving differently in the near and mid-IR.

\subsection{ISO-Oph 140: A source with a time-variable slope in the CMD}
\label{sect:isooph140asourcewithatimevariableslopeinthecmd}

\begin{figure*}
\plotone{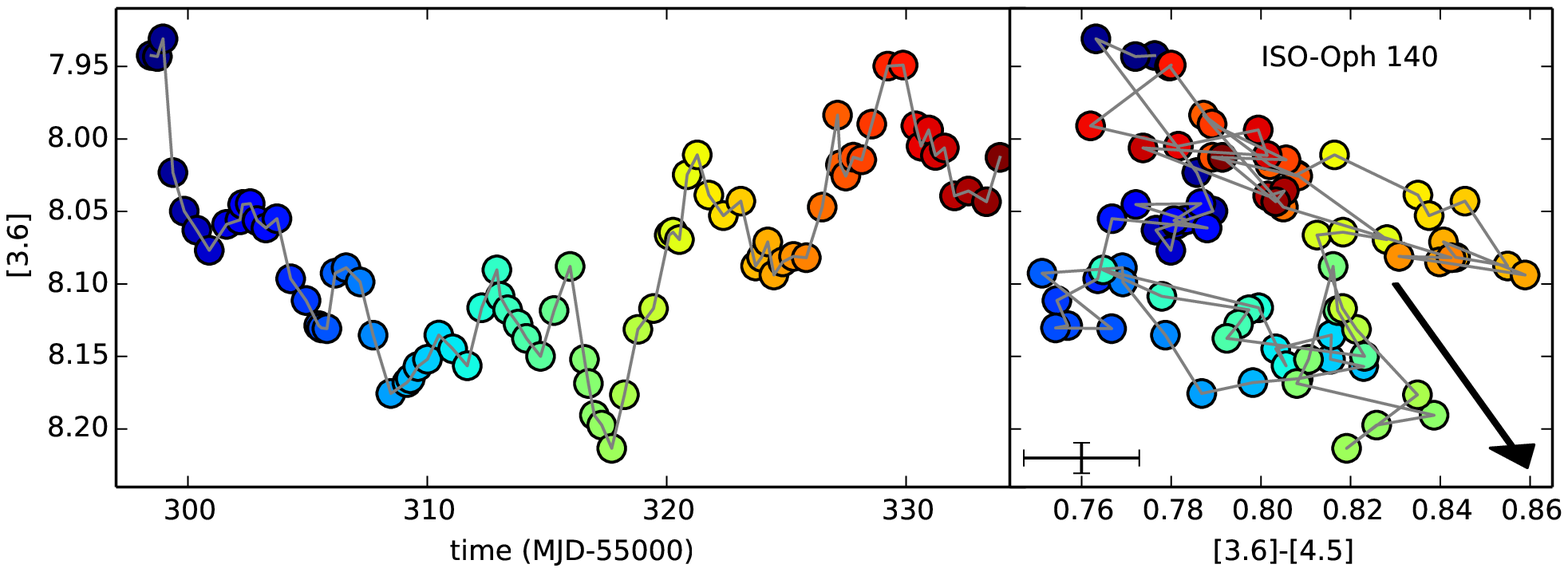}
\caption{Lightcurve (left) and CMD (right) for ISO-Oph 140 in the first observing visibility window. 
The color of the symbols in the CMD shows their position on the lightcurve.
Typical error bars are shown in the CMD. The arrow indicates a reddening of $A_K = 0.2$~mag \citep{2005ApJ...619..931I}.
\label{fig:cmdsquare}}
\end{figure*}
\object{ISO-Oph 140} is the only source with a statistical error on the slope in the CMD $>6^{\circ}$ where the color variability is not dominated by observational uncertainties, but where the slope varies substantially over time. Its lightcurve and CMD for the first visibility window are shown in Figure~\ref{fig:cmdsquare}. It has a class~II SED slope and is a low mass YSO \citep[spectral type M1;][]{1999ApJ...525..440L}. There is no single slope in the CMD; instead the source seems to switch between different modes. Initially, the luminosity drops sharply, while the color becomes slightly bluer. The behavior changes around MJD 55305. For the next 15 days, the source becomes redder and dimmer, but the slope is less steep than the ISM reddening law. The evolution is not monotonic, but a brightening (e.g.\ around MJD 55313) corresponds to a bluer color and the source reddens as it dims. At the end of this period, the brightness increases again sharply and the slope in the CMD is comparable to the period MJD 55295 to 55305: The source becomes noticeably redder with higher luminosity. In the last 10 days of the monitoring, the source again has a slope close to the ISM reddening law. Its luminosity continues to increase and the color bluens until it has a similar color and luminosity as it did at the beginning of the monitoring. This closes the circle in the CMD. This is the only source in our sample where we observe multiple changes of the reddening slope over time. In this case, variability similar to ISM reddening corresponds to slow changes in luminosity, while the other direction of the slope in the CMD is associated with faster luminosity changes.

\subsection{Large amplitude variability}
\label{sect:largeamplitudevariability}

Now we discuss the lightcurves with the largest changes in magnitude in our sample. WL~4, shown in Figure~\ref{fig:cmd}, has an outburst event consisting of a brightness increase by about 0.6~mag and bluening at the same time. The rise and fall take only a few days and the entire outburst lasts about a month.

[EDJ2009]~892 (shown in the same figure) presents a slow rise by about 0.2~mag in $4.5\;\mu$m and smooth decay over two years where the source gradually bluens (although more activity between visibility windows cannot be excluded). There is an indication of a slow and smooth increase in brightness again in the last observing visibility window, pointing to a recurring phenomenon. These timescales are much longer than the dynamical time in the inner disk, so they are presumably driven by disk phenomena that originate at larger radii.

\citet{2011ApJ...733...50M} identified a few such bursting or fading events in their analysis of YSOVAR data for the Orion Nebula Cluster (ONC) and they have also been observed in the optical \citep{2010ApJS..191..389C,2013ApJ...768...93F}. The physical cause is not known yet. \citet{2013ApJ...768...93F} find a few lightcurves that match theoretical predictions for short mass accretion events, but without simultaneous spectroscopy this is hard to prove. 

The best example presented here, WL~4, becomes bluer during the burst, but the opposite happens for WL~3, which otherwise has a similar lightcurve. Thus, it is likely that different mechanisms cause these burst events.

\subsection{Scenarios for color changes}
\label{sect:scenariosforcolorchanges}

The reddening sources in Figure~\ref{fig:cmdslope} have slopes that are roughly compatible with the slope of an interstellar reddening law. Small deviations can be explained by modifications of the dust-to-gas ratio or the grain size and composition. This is commonly observed in individual young stars \citep{2008A&A...481..735G} and star forming regions as a whole \citep{2010AJ....140..266W,2012AJ....144..101G}. 

The most prominent example of a class~II source where the extinction changes on timescales of days is \object{AA Tau}. In this case, the absorption dip is periodic and caused by an inner disk warp partially occulting the star \citep{1999A&A...349..619B,2007A&A...462L..41S}. 
%In the last two years, the behavior of AA~Tau changed. It is now much fainter. %\citet{2013arXiv1304.1487B} explain this with an optical reddening that increased by four %magnitudes, most likely due to an asymmetry in the disk that rotated into the line-of-sight. 
Many examples of comparable lightcurves have been found in the ONC by \citet{2011ApJ...733...50M} and in NGC~2264 by \citet{cody2014}. These authors call objects with absorption events in the lightcuve ``dippers''.

We observe reddening on timescales of months to years in about half of all sources with a well-defined slope in the CMD. An example of that is shown in the top right panel of Figure~\ref{fig:cmd}. We do not know the inclination of the objects in our sample, but if it is close to edge-on, this might be caused by an asymmetric disk similar to AA~Tau. The required warp or local change in scale height that lets some gas and dust protrude above the average disk height, could be caused by a low-mass star, brown dwarf or planetary mass object orbiting in the disk \citep[e.g.][]{2011ApJ...736...85U,2013LNP...861..201B}. Alternatively, vortices in the disk \citep[e.g.][]{2010A&A...513A..60L,2012ApJ...754...21L} or dust traps \citep{2013A&A...554A..95P} can also change the local scale height. 

Changes in the optical brightness and color of a YSO are often attributed to spots, either hot accretion spots or cold magnetic spots, rotating in and out of view or to time-variable extinction \citep[e.g.][]{2001AJ....121.3160C,2002AJ....124.1001C}. If the reddening increases and, at the same time, the visible fraction of the hot spot decreases, then the resulting slope in the CMD would be intermediate between ISM reddening and colorless.

Since spots on the star cannot explain sources that bluen as they dim, this observed phenomenon must be related to the disk. Several parameters of the inner disk could be time variable. The most obvious one is the accretion rate. However, the direct effect of increased accretion is a larger accretion spot, which would make the star bluer and brighter, not fainter. Also, \citet{2012PASP..124.1137F} find no relation between mid-IR lightcurves and spectroscopic accretion tracers in their limited sample. This indicates that the relevant parameter can change without affecting the accretion rate. In the mid-IR, we see the optically thick dust at the inner disk edge. 
%The physical temperature of the dust is determined by the distance to the central star and its
%chemical composition. The distance in turn is usually compatible with the stellar irradiation and the
%sublimation temperature. Dust that moves inward of the dust sublimation radius is heated and
%evaporates on timescales less than one orbital period \citep[see discussion in][]{2013AJ....145...66F}. If the irradiation is reduced, for example because the accretion rate drops and the accretion spot cools down, then the dust forms again and the inner disk radius moves inward. 
Because the disk is heated over its entire vertical height at the inner edge while only the disk surface absorbs radiation at larger radii, it can form a puffed-up inner rim that reaches above the usual disk scale height and casts a shadow on the remaining disk, reducing the luminosity at longer wavelengths \citep[e.g.][]{2008ApJ...682..548W,2010ApJ...717..441E,2011ApJ...728...49E}. If this rim grows, it intercepts more stellar light. Thus, the emission from the inner wall increases and the emission from the outer disk decreases, which results in bluer colors. Since the total emitting area also decreases, the source can become bluer and fainter. 
%However, since the emission at $3.6\mu$m and $4.5\mu$m is formed close to the sublimation radius, this requires changes at those radii, perhaps in the shape and not only the height of the inner wall. %Time variable self-shadowing has been observed with direct imaging in \object{HD 163296} \citep{2008ApJ...682..548W} and \citet{2010ApJ...717..441E} present a series of \emph{Spitzer} spectra of T~Tauri stars where the spectrum can be modeled taking into account the shadows cast by the inner disk rim. In a time series of spectra, those models explain a decreasing far-IR ($>20\mu$m) flux that coincides with an increase in emission from the inner wall \citep{2011ApJ...728...49E}.
\citet{2013AAS...22125610K} also present radiative transfer models with hot spots and warped disks that reproduce features of the observed lightcurves.

Alternatively, \citet{2013AJ....145...66F} suggested a non-axisymmetric model. If the emission from the stellar photosphere is stronger at some longitudes due to a strong accretion spot or a spot from magnetic activity on the star, it irradiates the inner disk like a searchlight beam and causes the dust at this longitude to retreat. The result is a disk where the inner rim is broken at one (or several) positions. This would reduce the emission in $3.6\,\mu$m, but cause more energy to be emitted at a larger radius, where the dust is cooler on average and thus emits at longer wavelengths such as $4.5\,\mu$m. In this way, the source can become redder as it brightens. 
%and bluer when the searchlight beam stops and the dust at the inner rim forms again at all longitudes. In this scenario, the time variability of the mid-IR emission should follow the accretion rate or the pattern of the searchlight and, if this is emitted by the accretion spot we expect to find the timescale of the stellar rotation period (if the change in brightness is a geometrical effect with the line-of-sight). With few exceptions, we see the lightcurves changing non-periodically with timescales of just a few days, faster than typical rotation periods of YSOs \citep{1989A&A...211...99B,2008A&A...479..827G}, and a relation between the mid-IR luminosity and the accretion rate is not seen in observations either.

There is one group of YSOs, the so-called UX Ori (UXOr) variables where several members have been observed to bluen when they fade in optical observations \citep{1990A&A...236..155B,1998ARA&A..36..233W}. This can be explained by a larger fraction of scattered light that contributes to the observed SED in their faint state. A similar scenario for the YSOs discussed here requires, first, scattering by relatively large grains ($>1$~mm), because only for large grains radiation at $3.6\;\mu$m will be scattered more than radiation at $4.5\;\mu$m \citep{1984ApJ...285...89D,2013A&A...559A..60A}. Grains this large will be present in the disk, but not at all radii and all disk heights. Second, for this mechanism to work, a large fraction of the IR radiation must be scattered light and not intrinsic emission. In contrast, many bluening sources are class~I to flat-spectrum sources, where the IR luminosity is comparable to the total stellar irradiation.

None of the simple models presented so far offers a convincing explanation of the observed lightcurves and CMDs. However, detailed simulations that take into account turbulent transport processes in the disk and different dust species with a complete chemical network have not yet been performed. The real situation is likely to be much more complex than sketched above. Since different dust species form and sublimate with different speeds and at different temperatures, time variable irradiation can actually cause a very complex mixture of species with different opacities, which will not react linearly to increased irradiation. Such a complex network might cause a delayed and non-linear response of the disk, which masks the underlying relation between accretion and disk emission.

\subsection{Variability and the host star}
\label{sect:variabilityandthehoststar}

The characteristics of the time variability in the lightcurves could be related to the spectral type of the central star as hotter stars irradiate the disk with harder spectra. We compared the 90\% quantiles of the observed $4.5\,\mu$m magnitudes for those sources in our sample that have spectroscopically determined spectral types from \citet{2011AJ....142..140E}. We use the relation between spectral type and $T_{\textrm{eff}}$ from \citet{2013arXiv1307.2657P}. This gives values for ten objects in our sample and we do not find a correlation in this set. Similarly, \citet{cody2014} do not identify any trends in infarared variability
versus effective temperature among few Myr old NGC 2264 stars.
In apparent contrast, \citet{2013AJ....145...66F} find that the variability in [3.6] and [4.5] increases with increasing $T_{\textrm{eff}}$, but they show that this trend probably does not reflect a change in disk properties, but is instead due to the lower relative contribution of the photosphere compared to the disk for hotter stars.

\subsection{X-ray emission and variability}
\label{sect:xrayemissionandvariability}

In this section, we analyze the subsample of sources with X-ray counterparts. Since the subsample is not selected for its IR properties, it provides a clean sample to calculate the variability fraction. There are 31 sources with detected X-ray emission, of which 20 are variable in the IR. Again, we find high variability fractions in class~I (6/7), F (3/4) and II (11/15), while none of the four class~III sources is variable. The sample size is small, but this is consistent with our analysis of the IR selected sample in section~\ref{sect:evolutionarytrendsofthevariability}. 

We searched for correlations between the parameters of the X-ray emission (median energy, fitted temperature and absorbing column density) and the amplitude of the mid-IR variability. No such correlation is apparent within each SED class. For a given absorbing column density, soft X-ray emission is more strongly absorbed than hard X-ray emission. Thus, the observed X-ray spectrum from more embedded sources always appears harder. Since class~II sources have more circumstellar matter than class~III sources, it is not surprising that they appear harder. \citet{2001ApJ...557..747I} already discuss this for the L1688 X-ray data; Figure~\ref{fig:probvarSED} and \ref{fig:meanampSED} show that class~II sources are more variable than class~III sources. We see no additional effect of the X-ray emission on the variability characteristics in the mid-IR.

\section{Summary}
\label{sect:summary}

We present \emph{Spitzer} observations of YSOs in the star forming region L1688. Observations were taken in four visibility windows 
in Spring and Fall of 2010 and 2011, with about 70 observations in Spring 2010 and about ten observations in the remaining visibility windows. The cadence of 
the observations is non-uniform to avoid bias in the period detection. 
Our sample consists of 882 sources with lightcurves in IRAC1 and IRAC2 with at least 5 datapoints.
Of those 882 sources, we classify 70 sources as variable using the
Stetson test, the $\chi^2$ test and the Lomb-Scagle periodogram. 
The faintest sources in the sample have $\sim16^{\mathrm{th}}$~mag, but naturally the measurement uncertainties are larger for fainter sources. 
The algorithms detect variability if 
the amplitude is larger than $\sim0.05$~mag for a source of $14^{\mathrm{th}}$~mag.

We define a sample of cluster members, including sources with an IR excess due to a circumstellar disk or with X-ray emission.
For cluster members, there is a clear correlation between evolutionary status and IR variability. More embedded sources are more often
detected to be variable, and they have on average larger variability amplitudes. Overall, the data are consistent with the idea that all
YSOs are variable at $3.6\,\mu$m and $4.5\,\mu$m, and we thus propose that all variable sources in our sample are members of
the L1688 star forming region.

Qualitatively different morphological types of lightcurves can be distinguished: 14 lightcurves are detected
to be periodic; beyond that, we find quasi-periodic lightcurves, where the variability has an apparent timescale, but is not 
regular enough to be detected as periodic; aperiodic lightcurves without a preferred timescale; and long-term
variable lightcurves where variability is apparent, but no periodicity or time-scale can be determined within the observational
window. In addition, there are lightcurves with short, non-repeating bursts or dips. 

Roughly half of all sources become redder when they are fainter; the other half becomes bluer. The reddening values of the first group
are compatible with ISM reddenig. The color changes in the second group require variability in the inner disk structure as proposed
by \citet{2010ApJ...717..441E} or \citet{2013AJ....145...66F}.

\acknowledgements
This work is based on observations made with the Spitzer Space Telescope, which is operated by the Jet Propulsion Laboratory, California Institute of Technology under a contract with NASA. Support for this work was provided by NASA through an award issued by JPL/Caltech.
This research made use of Astropy, a community-developed core Python package for Astronomy \citep{2013arXiv1307.6212T}. This research has made use of the SIMBAD database and the VizieR catalogue access tool \citep{2000A&AS..143...23O}, both operated at CDS, Strasbourg, France and of data products from the Two Micron All Sky Survey, which is a joint project of the University of Massachusetts and the Infrared
Processing and Analysis Center/California Institute of Technology, funded by the National
Aeronautics and Space Administration and the National Science Foundation. H.M.G. acknowledges Spitzer grant 1490851.
H.Y.A.M.\ and P.P.\ acknowledge support by the IPAC Visiting Graduate Fellowship program at Caltech/IPAC.
P.P.\ also acknowledges the JPL Research and Technology Development and Exoplanet Exploration programs.
RAG gratefully acknowledges funding support from NASA ADAP grants NNX11AD14G and NNX13AF08G and Caltech/JPL awards 1373081, 1424329, and 1440160 in support of Spitzer Space Telescope observing programs. S.J.W. was supported by NASA contract NAS8-03060.

{\it Facilities:} \facility{Spitzer} \facility{Chandra}


\appendix

\section{Remarks about cross-matching individual sources}
\label{sect:remarksaboutcrossmatchingindividualsources}

\begin{description}
\item[\object{2MASS J16272802-2439335}]
We identify 2MASS J16272802-2439335 with \object{[AMD2002] J162728-243934A} because they match within the positional accuracy and the observed flux densities fit nicely together in a SED.
\item[\object{2MASS J16273288-2428116}]
After a visual inspection of the 2MASS images, which form the basis of \citet{2010ApJ...719..550M}, we also identify 2MASS J16273288-2428116 with \object{[MPK2010b] 1307}.
\item[\object{[GPJ2008] Source 3}]
Based on the position and the expected shape of the SED for a YSO, we also identify [GPJ2008] Source 3 with \object{[SSG2006] MMS002}.
\item[\object{[AMD2002] J162724-242850}]
We propose that [AMD2002] J162724-242850 and \object{[MPK2010b] 4077} are the same source, because their positions match and the fluxes form an SED that is fully consistent with a YSO in L1688.
\item[\object{SSTc2d J162621.7-242250} and \object{[GMM2009] Oph L1688 3}]
\label{sect:misidentified}
In comparison with \citet{2009ApJS..184...18G}, we find that \object{SSTc2d J162621.7-242250} is erroneously identified with source Oph~L1688~3 in SIMBAD. The distance between the positions on the sky is 1\farcs3, much larger than the positional uncertainty. Therefore, the name \object{[GMM2009] Oph L1688 3} is not used in our source table.
\item[YLW 16]
In one case, the SIMBAD database does contain a reference to a multiple source (\object{YLW 16}), that is resolved in the IR. We associate our lightcurves with \object{YLW 16A}, which is the dominant component.

\item[\object{ISO-Oph 152}]
The UKIDSS $K$-band image shows at least two sources within 1\arcsec{} of ISO-Oph~152 with partially overlapping point-spread functions (sourceIDs 442426789990 and 442426789990). The position of the second source is uncertain and fitted differently in the other bands (where it is called source ID 442426789991). All of this is not resolved in the IRAC data; thus, in each band, we assign the brightest magnitude of either of those sources to ISO-Oph~152.

\end{description}

\section{Sources that are YSOs in G09, but not listed as members in Wilking et al. (2008)}
\label{sect:sourcesthatareysosing09butnotlistedasmembersinwilkingetal2008}

\label{sect:g09notwil}
\begin{description}
\item[\object{CFHTWIR-Oph 29}] This source detected in several near-IR observations and is known to be variable in that wavelength range \citep{2008A&A...485..155A}.
\item[\object{LFAM 4}]. This source is part of the triple system that is well resolved in sub-mm
   observations. Based on its SED stretching out to 6~cm, this is a class~I source, which matches the 
    classification we derive from the IR data. Most likely, this souce contributes to the seizable outflows
    observed from the triple system, which would further confirm its youth and thus its status as a
    cluster member \citep{2013ApJ...764L..15M}.
\item[\object{[GMM2009] Oph L1688 115}] This source is detected at $1.6\;\mu$m, but not at $1.1\;\mu$m by
    \citet{2002ApJ...566..993A} but no further information is available in the literature.
\item[\object{[GMM2009] Oph L1688 30}] In \citet{2008ApJ...683..822J}, this objects is listed as a YSO and potential member of L1688 based on the G09 SED classification and a match to a SCUBA source at $850\;\mu$m. However, the next peak of the SCUBA flux is 29\arcsec{} from the position of [GMM2009] Oph L1688 30, just below the maximum distance for a match that is accepted in that work.
\item[\object{[EDJ2009] 824}] This source is also listed as a YSO with a SCUBA flux peak 20\arcsec{} away \citep{2008ApJ...683..822J}.
\item[\object{SSTc2d J162621.7-242250}] We find that this source is mismatched in SIMBAD (see Section~\ref{sect:misidentified}). There are several radio observations in molecular lines and in the continuum \citep{2007ApJ...671.1800A,2008ApJ...683..822J,2009A&A...498..167V} which indicate substantial circumstellar material and thus confirm the status as a YSO, but --if SSTc2d J162621.7-242250 and [GMM2009] Oph L1688 3 are different sources-- either of them could be the source of the radio signatures.

\end{description}

\end{document}